\documentclass{aa}
\usepackage{natbib}
\usepackage[version=4]{mhchem}
\usepackage[normalem]{ulem}
\usepackage[dvipsnames]{xcolor}
\usepackage{caption}
\usepackage{placeins}
\usepackage{float}
\usepackage{textcomp}
\bibpunct{(}{)}{;}{a}{}{,} 
\usepackage{graphicx}
\usepackage{txfonts}
\usepackage{amsmath}	
\usepackage{amssymb}	
\begin{document} 

   \title{Observability of evaporating lava worlds}

   \author{M. Zilinskas
          \inst{1}
          \and
          C.P.A. van Buchem\inst{1} \and
          Y. Miguel\inst{1,2} \and
          A. Louca\inst{1} \and
          R. Lupu\inst{3} \and
          S. Zieba\inst{4,1} \and
          W. van Westrenen\inst{5}.
          }

   \institute{Leiden Observatory, Leiden University, Niels Bohrweg 2, 2333CA Leiden, the Netherlands
   \and
                SRON Netherlands Institute for Space Research , Niels Bohrweg 4, 2333 CA Leiden, the Netherlands
    \and
              Eureka Scientific Inc, Oakland, CA 94602
               \and
                Max-Planck-Institut f\"ur Astronomie, K\"onigstuhl 17, D-69117 Heidelberg, Germany
               \and
               Department of Earth Sciences, Vrije Universiteit Amsterdam, De Boelelaan 1085, 1081 HV Amsterdam, the Netherlands
                \\
            \\
              \email{zilinskas@strw.leidenuniv.nl}
             }

   \date{Received June XX, 2021; accepted July XX, 2021}

 
  \abstract
  {Lava worlds belong to a class of short orbital period planets reaching dayside temperatures high enough to melt their silicate crust. Theory predicts that the resulting lava oceans outgas their volatile components, attaining equilibrium with the overlying vapour. This creates a tenuous, silicate-rich atmosphere that may be confined to the permanent dayside of the planet. The James Webb Space Telescope (JWST) will provide the much needed sensitivity and spectral coverage to characterise these worlds. In this paper, we assess the observability of characterisable spectral features by self-consistently modelling silicate atmospheres for all the currently confirmed targets having sufficient sub-stellar temperatures (> 1500 K). To achieve this we used outgassed equilibrium chemistry and radiative transfer methods to compute temperature-pressure profiles, atmospheric chemical compositions, and emission spectra. We explore varying melt compositions, free of highly volatile elements, accounting for possible atmospheric evolution. Our models include a large number of neutral and ionic species, as well as all up-to-date opacities. The results indicate that \ce{SiO} and \ce{SiO2} infrared features are the best unique identifiers of silicate atmospheres, which are detectable using the MIRI instrument of JWST. Detection of these two species in emission would allow for strong constraints on the atmospheric thermal structure and possibly the composition of the melt. We also propose that certain species, for example \ce{TiO}, may be directly tied to different classes of melts, possibly revealing surface and interior dynamics. Currently, there are nearly a dozen confirmed lava planets ideal for characterisation of silicate atmospheres using JWST, with two of these already accepted for the initial General Observers programme.
  }
  


   \maketitle

\section{Introduction}
\begin{figure}
        \centering
    	\includegraphics[width=0.5\textwidth]{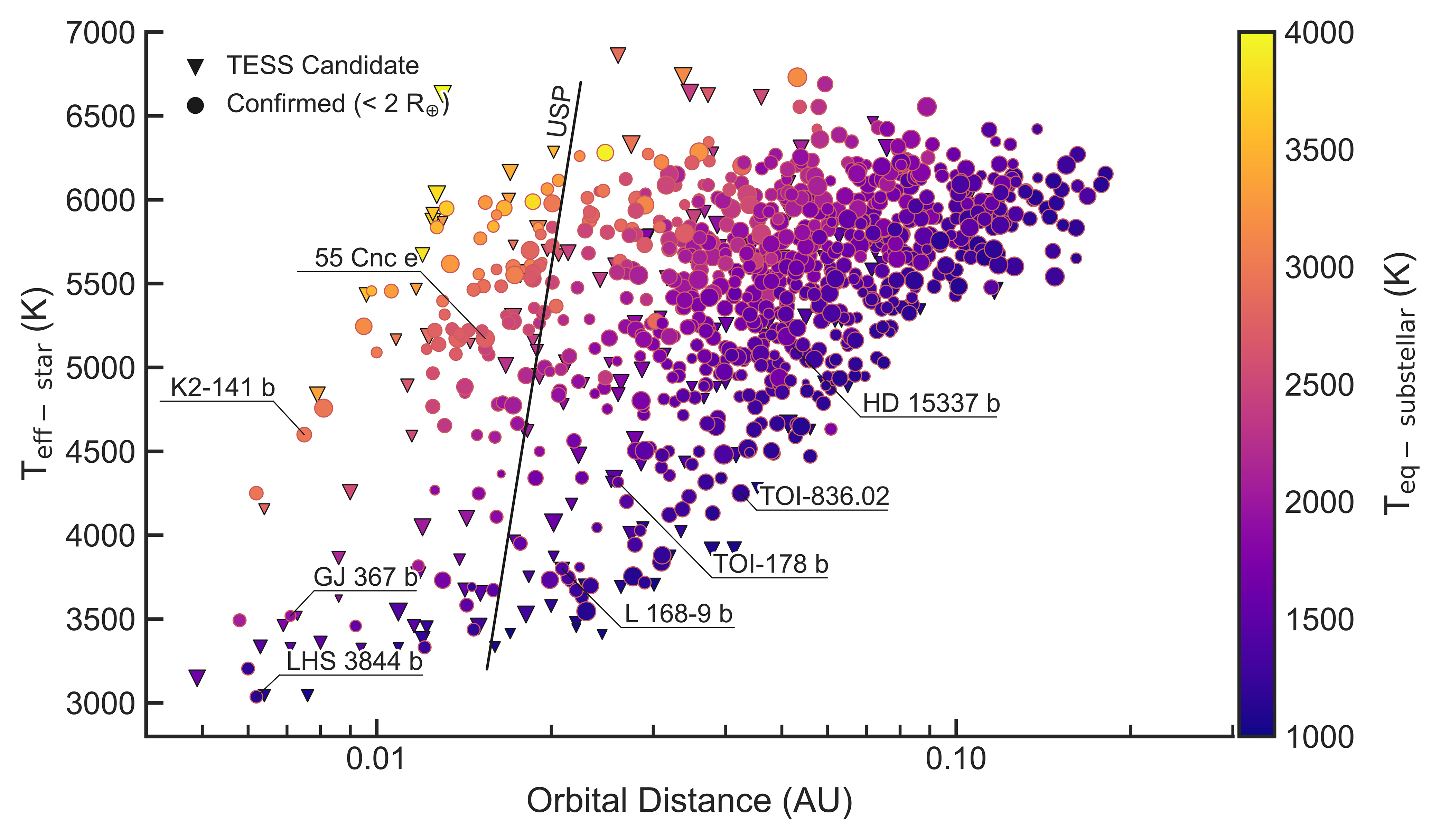}
        \caption{Confirmed transiting short period exoplanets (including current TESS candidates) with radii < 2 R${_\oplus}$ and equilibrium sub-stellar temperatures > 1000 K. Planets to the left of the ultra-short period (USP) line have orbital periods < 1 day. Named planets are approved for the JWST Cycle 1 GO programme. The colour indicates the theoretical sub-stellar temperature of the planet. The marker size is proportional to the radius of the planet. Over 800 confirmed planets and more than 150 TESS candidates are displayed. The complete table of all shown planets with their parameters can be downloaded as supplementary material.}
        \label{fig:Planets}
    \end{figure}

   The James Webb Space Telescope (JWST) Cycle 1 General Observers (GO) programme has approved observations of 17 rocky exoplanets\footnote{Approved JWST Cycle 1 GO rocky planets. Highly irradiated: K2-141 b, 55 Cnc e, GJ 367 b, TOI-178 b, HD 15337 b, L 168-9 b, TOI-836.02, and LHS 3844 b. Others: GJ 486 b, TRAPPIST-1 b, c, g, LHS 1140 b, TOI-776 b, GJ 357 b,  LTT 1445 A b and L 98-59 c.}, with eight most irradiated on the list having sub-stellar temperatures large enough to sustain a molten surface crust. Close-in rocky planets, due to their short orbital periods, are readily found by survey missions like the Transiting Exoplanet Survey Satellite \citep[TESS;][]{Ricker_2014}. Because of their large day-side temperatures, which allow for formation of silicate atmospheres, lava planets are particularly interesting targets for follow up characterisation with infrared spectroscopy. Figure \ref{fig:Planets} showcases all currently confirmed < 2 R${_\oplus}$ (along with TESS candidates) transiting exoplanets with equilibrium sub-stellar temperatures above 1000 K (assuming zero albedo). 
 
   Extensive research has been conducted on silicate atmosphere formation and the possibility of their characterisation. Models by \citet{Schaefer_2009} showed that silicate atmospheres would be composed primarily of \ce{Na}, \ce{O2}, and \ce{SiO}. Based on their work, \citet{Miguel_2011} studied the entire population of Kepler super-Earths and found that these atmospheres are likely to be diverse in their composition, which is strongly dependant on the given composition of the melt and the irradiation of the planet. By additionally considering radiative-transfer, \citet{Ito_2015} investigated emission properties of irradiated super-Earths and discovered that their atmospheric temperature structure may be completely inverted. Their work also points to the possibility of characterising lava planets through the emission features caused solely by the presence of \ce{SiO}. 
   
   It has also been proposed that because of their tenuous nature, silicate atmospheres may be prone to temporal evolution \citep{Schaefer_2009, Kite_2016}, as most volatile constituents could be driven to condense out in the cooler regions of the planet or even escape the atmosphere entirely. Because the atmosphere is essentially coupled to the underlying magma, atmospheric evolution could irreversibly change the composition of the melt. The effectiveness of this, however, is strongly dependant on the rapidity of material recycling in the interior \citep{Kite_2016}. Following this, \citet{Nguyen_2020} and \citet{Ito_2021} modelled evaporation-driven hydrodynamics with radiative-transfer and found that atmospheres dominated by \ce{SiO} are much more likely to be stable against circulation and escape effects compared to ones dominated purely by \ce{Na}. The current consensus is that silicate atmospheres may lack volatile \ce{Na} and instead would be largely dominated by \ce{SiO} or \ce{SiO2}, both of which could be characterised through emission features.
   
   Following the approach from \citet{Miguel_2011} and \citet{Ito_2015} and motivated by the possible evaporation evolution \citep{Schaefer_2009,Kite_2016,Nguyen_2020}, we extensively modelled outgassed equilibrium chemistry and the atmospheric structure in order to investigate JWST observability for the shown short period rocky planets (Fig. \ref{fig:Planets}). We put constraints on the current target list and indicate possible atmospheric features that could help characterise lava planets through low resolution spectroscopy. Our models are self-consistent, accounting for the changes in thermal structure caused by the chemistry, and were done with the inclusion of all up-to-date opacities that could potentially shape silicate atmospheres.
   
   This paper is organised as follows. In Section \ref{sec:Methods} we describe our approach to modelling silicate atmospheres, including the use of outgassed equilibrium chemistry and radiative transfer methods. We present our findings for non-evolved atmospheres in Section \ref{sec:ResultsI}. For evolved atmospheres, the results are outlined in \ref{sec:ResultsII}. Section \ref{sec:ResultsIII} highlights the results of models with different initial melt compositions. Overall implications and observability of silicate atmospheres are discussed in detail in Section \ref{sec:Discussion}. Finally, we conclude by summarising our key findings in Section \ref{sec:Conclusions}.

\section{Methods}
\label{sec:Methods}
\subsection{Outgassed chemistry}

Sustained planetary surface magma will preferentially outgas its most volatile components and eventually equilibrate itself with the overlying vapour, forming a silicate atmosphere. We determine the total atmospheric elemental budget by using the published results of the melt-vapour equilibrium code \texttt{MAGMA}, which has been used to calculate outgassing on solar system bodies and exoplanets \citep{Fegley_1987,Schaefer_2004,Schaefer_2009,Miguel_2011,Kite_2016}. Its output is determined by the given initial composition of the melt, surface temperature and evaporation percentage. The involved chemistry considers the oxides of: \ce{Al}, \ce{Ca}, \ce{Fe}, \ce{K}, \ce{Mg}, \ce{Na}, \ce{Si} and \ce{Ti}. We make the assumption that due intense irradiation from the host star, highly volatile elements (H, C, N, S) are completely void from the atmosphere. In such a case the atmospheric composition is controlled only by the continuous evaporation of the underlying melt and not by past accretion of volatiles \citep{Schaefer_2009,Kite_2016}.

Given the diverse population of stars, it is likely that melt compositions are highly varied. Approximations can be made from the particular host star \citep{Brugman_2021}, this is however beyond the scope of this study. Bulk Silicate Earth (BSE) has been widely used in research of silicate atmospheres \citep{Schaefer_2009,Miguel_2011,Ito_2015,Kite_2016}. We adopt BSE as the main composition, but also briefly investigate a selection of other possible, silicate based melts. All used compositions are listed in Table \ref{table:melts}.

In our models the surface temperature is solved for the sub-stellar point, which is continuously adjusted using radiative-transfer methods that couple the surface and the overlying atmospheric layer (described in Section \ref{sec:MethodsT}). This is process is done until convergence between the outgassed chemistry and thermal structure is achieved.

\citet{Kite_2016} has shown that for atmospheres with sub-stellar temperatures > 2400 K, atmospheric evolution is faster than surface-interior exchange, implying that continuous evaporation would rapidly drive the surface magma composition away from its initial values. While we do not directly simulate loss of material from the atmosphere through condensation or escape, we mimic the process using continuous removal of the most volatile components from the system. As described in \citet{Schaefer_2004}, this is done in a step-wise manner, until a set percentage of total melt mass is lost. We explore continuous removal of up to 60 percent mass of the given melt composition.

\begin{table*}
 \captionsetup{justification=centering}
 \caption[]{Initial melt compositions}
 \label{table:melts}
 \centering
 \begin{tabular}{llllllllll}
    & \ce{SiO2} & \ce{MgO} & \ce{FeO} & \ce{Fe2O3} & \ce{Al2O3} & \ce{CaO} & \ce{Na2O} & \ce{TiO2} & \ce{K2O}\\
  \hline
  \hline
  \\
  BSE$^a$ & 45.97 & 36.66 & 8.24 & 0.0 & 4.77 & 3.78 & 0.35 & 0.18 & 0.04\\
  Continental$^b$ & 62.93 & 3.79 & 5.78 & 0.0 & 15.45 & 5.63 & 3.27 & 0.7 & 2.45\\
  Oceanic$^c$ & 50.36 & 7.61 & 9.56 & 0.0 & 15.85 & 12.24 & 2.77 & 1.48 & 0.13\\
  Komatiite$^d$ & 47.1 & 29.6 & 0.0 & 12.8 & 4.04 & 5.44 & 0.46 & 0.24 & 0.09\\
  Mercury$^e$ & 47.1 & 33.7 & 3.75 & 0.0 & 6.41 & 5.25 & 0.08 & 0.33 & 0.02\\
  \\
  \hline
  \multicolumn{10}{p{13.4cm}}{\footnotesize$^{a}$Bulk Silicate Earth \citep{Schaefer_2009}; \footnotesize$^{b}$Continental crust \citep{Wedepohl_1995};  \footnotesize$^{c}$Oceanic \citep{Klein_2003}; \footnotesize$^{d}$Komattite \citep{Schaefer_2004}; \footnotesize$^{e}$ Mercury \citep{Morgan_1980}}
\\
 \end{tabular}
\end{table*}

With elemental abundances established, atmospheric chemistry is solved using the thermochemical equilibrium code \texttt{FASTCHEM}\footnote{https://github.com/exoclime/FastChem} \citep{Stock_2018} in accordance with the thermal structure of the atmosphere. We consider the following neutral components: \ce{Al}, \ce{AlO}, \ce{AlO2}, \ce{Al2O}, \ce{Al2O2}, \ce{Ca}, \ce{CaO}, \ce{Fe}, \ce{FeO}, \ce{FeO2}, \ce{K}, \ce{KO}, \ce{K2}, \ce{K2O}, \ce{Mg}, \ce{MgO}, \ce{Na}, \ce{NaO}, \ce{Na2}, \ce{Na2O}, \ce{O}, \ce{O^1}, \ce{O2}, \ce{O2^1}, \ce{O3}, \ce{Si}, \ce{SiO}, \ce{SiO2}, \ce{Si2}, \ce{Ti}, \ce{TiO}, \ce{TiO2}; as well as ions: \ce{Al+}, \ce{Ca+}, \ce{Fe+}, \ce{K+}, \ce{Mg+}, \ce{Na+}, \ce{O+}, \ce{O-}, \ce{O2+}, \ce{O2-}, \ce{O3-}, \ce{Si+}, \ce{Ti+}, \ce{TiO+}. The thermal data is compiled from Burcat NASA thermodynamics polynomial database\footnote{http://garfield.chem.elte.hu/Burcat/burcat.html}. We note that our inclusion of various components is only limited by the availability of thermodynamical data.

\subsection{Temperature structure}
\label{sec:MethodsT}


We solve for the temperature structure using the radiative-transfer code \texttt{HELIOS}\footnote{https://github.com/exoclime/HELIOS} \citep{Malik_2017,Malik_2019}. Profiles are converged to radiative-convective equilibrium consistently with the chemistry (convection has negligible effect for silicate atmospheres), assuming no surface albedo. The surface temperature itself is affected by the overlying atmospheric and is thus calculated according to the propagating flux. The exact implementation of the surface layer for rocky planets in \texttt{HELIOS} can be found in \citet[][Appendix A]{Malik_2019b}.

We consider 25 different opacity sources that could potentially affect the temperature structure, which the list of can be seen in Fig. \ref{fig:Opacity}. Atomic opacities (excluding \ce{Na}) are calculated with the opacity calculator \texttt{HELIOS-K}\footnote{https://github.com/exoclime/HELIOS-K} \citep{Grimm_2015,Grimm_2021}, for which we use the Kurucz \citep{Kurucz_1992} line list database. Molecular opacities (excluding \ce{SiO}) are sourced from the DACE\footnote{https://dace.unige.ch/} database and/or computed with \texttt{HELIOS-K}. For both, atoms and molecules, we use Voigt fitting profiles. Following the methods from \citet{Grimm_2015,Grimm_2021}, we use a cutting length of 100 cm$^{-1}$ for molecular species and no cutting for atomic species (excluding \ce{Na} and \ce{SiO}).

Since \ce{Na} and \ce{SiO} are likely to be major components of silicate atmospheres, we compute these opacities separately. \ce{Na} opacity is calculated using the Vienna Atomic Line Database (VALD3) line list \citep{Ryab_2015}, approximated using Voigt profiles for all but the 0.6 \textmu m doublet lines. For the doublet, standard Lorentzian fitting is inadequate. Instead, it is fitted using the unified line-shape theory of \citet{Rossi_1985,Allard_2007a,Allard_2007b}. Our \ce{SiO} opacity is constructed using the \citet{Barton_2013} line list for the ground state transitions in the infrared and with the  \citet{Kurucz_1992} line list for the A-X and E-X shortwave wavelength transitions.

We emphasise that these opacities are modelled with parameters applicable to atmospheres of stars and gas giants, where the background is dominated by \ce{H2} and \ce{He} \citep{Marley_2021}. There are no currently existing equivalent theories for silicate atmospheres. While we do find that inclusion such of pressure broadening may have non-negligible effects on the temperature structure, these are relatively minor for silicate atmospheres, typically resulting in less than $\Delta T$ = 100 K close to the surface, but otherwise nearly identical temperature-pressure profiles. To minimise the possibility of overestimation we take opacity values at lowest available pressures, resulting in Doppler-dominated solutions. Use of this assumption is one of the limiting factors of the current work, emphasising the need for future development of broadening theories for arbitrary atmospheres. The complete parameters and sources of all the opacities can be found in the Appendix Table \ref{table:opacities}.

The last thing to account for is stellar irradiation. For cooler stars (e.g., M dwarfs), blackbody stellar spectra inaccurately approximates shortwave energy flux. When dealing with high temperatures, shortwave absorption becomes an important factor in shaping the thermal structure of an atmosphere. We thus construct stellar spectra\footnote{All individual spectra are available for download as supplementary material https://github.com/zmantas/LavaPlanets.} for each of the individual stars by combining PHOENIX models \citep{Husser_2013} with scaled shortwave MUSCLES spectra \citep{France_2016,Youngblood_2016,Loyd_2016}. This method is only applied to low mass stars, with temperatures covered in the MUSCLES database. For any hotter stars we scale the shortwave VPL solar spectrum\footnote{http://depts.washington.edu/naivpl/content/spectral-databases-and-tools}, in combination with PHOENIX models. All opacities and stellar spectra are sampled at a resolution of $\lambda/\Delta\lambda = 2000$, accounting for wavelengths between 0.1 and 200 \textmu m.

\subsection{Synthetic emission spectra}
\label{sec:MethodsE}

Dayside confinement makes silicate atmospheres poor candidates for low resolution transmission spectroscopy. However, the large temperatures make secondary eclipse observations much more feasible. We generate emission spectra using the radiative-transfer code \texttt{petitRADTRANS}\footnote{http://gitlab.com/mauricemolli/petitRADTRANS} \citep{Molliere_2019,Molliere_2020}. Using the same opacities described in Section \ref{sec:MethodsT} we calculate the spectra at a resolution $\lambda/\Delta\lambda = 1000$ for wavelengths between 0.3 - 28 \textmu m. In many cases the spectra are convolved to a lower resolution for better readability. The explored wavelength range encompasses the coverage of all instruments of JWST.

For notable targets we assess JWST noise levels using \texttt{PANDEXO}\footnote{https://exoctk.stsci.edu/pandexo/} \citet{Batalha_2017}, which is built on the Pandeia\footnote{http://jwst.etc.stsci.edu} engine. We only simulate MIRI Low Resolution Spectroscopy (MIRI LRS with $\lambda/\Delta\lambda \approx 100$) in slitless mode, as it is likely to be the most suitable mode for characterisation of silicate atmospheres. The wavelengths covered by the instrument are 5 - 12 \textmu m.

\section{Results}
\label{sec:Results}

\subsection{Bulk silicate earth atmospheres}
\label{sec:ResultsI}
\subsubsection{Opacities \& Chemistry}

The thermal structure in an atmosphere is shaped by wavelength dependant absorption of the incoming stellar radiation. Silicate atmospheres possess numerous different species that are good absorbers in the UV/Visible (shortwave) and infrared (IR) wavelengths. Generally, if the opacity at short wavelengths exceeds that of IR, the temperature of the absorbing region increases, causing an inversion to occur \citep{Gandhi_2019,Malik_2019}. In cases with no shortwave absorbers, efficient cooling through IR will cause the temperature to decrease with increasing altitude.

\begin{figure*}[!h]
    \centering
	\includegraphics[width=1.0\textwidth]{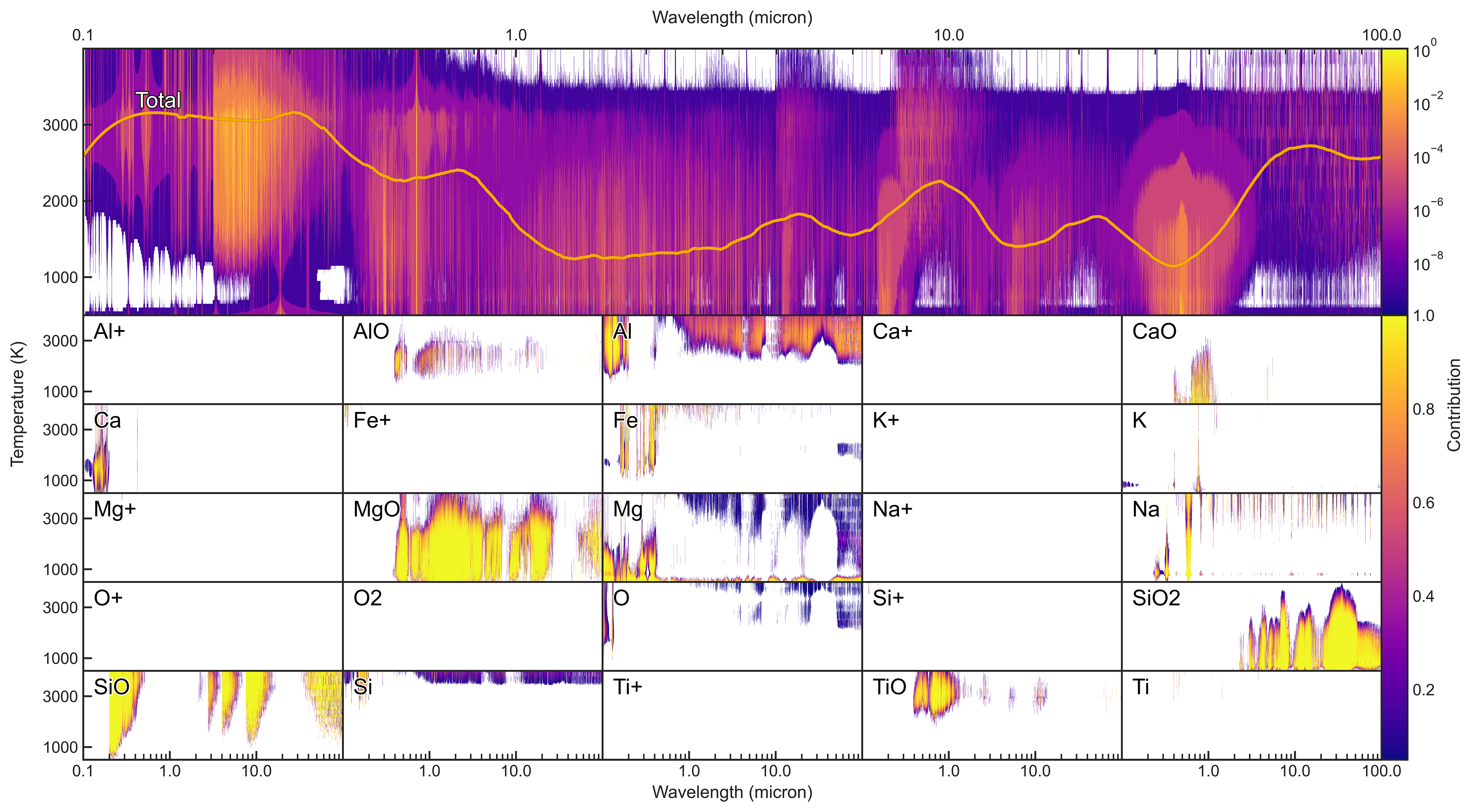}
    \caption{Sampled opacities of a silicate atmosphere. Shown for a typical Bulk Silicate Earth melt composition with a surface temperature of $\sim$2700 K and atmospheric pressure of $10^{-4}$ bar. The individual species indicate contributed absorption to the total opacity, with the minimum cutoff threshold of 5\%. The upper panel shows temperature-specific absorption contribution relative to the maximum opacity for all wavelengths, with a minimum threshold of $10^{-10}$. The curve is fitted to indicate the maximum, wavelength-specific opacity for all temperatures. Note that the labelled temperatures are atmospheric, which represent the temperature of the melt.}
    \label{fig:Opacity}
\end{figure*}

Figure \ref{fig:Opacity} shows example opacities of the 25 absorbing species investigated in this study. All values are weighted to abundances of a non-evolved stage of an atmosphere, formed from a BSE melt. Values are shown for a pressure of $10^{-4}$ bar, a typical region where the photosphere begins to take shape. For each given temperature and wavelength, the opacity contribution is calculated relative to the total added opacity of all present species. Different species will dominate distinct wavelength regions, allowing us to easily track their effect on the thermal structure and emission spectrum.

Most ions, even though increasingly abundant in hot atmospheres, can be dismissed as potential absorbers. Their opacities are significantly lower in value compared to their neutral counterparts. There are exceptions to this, notably \ce{Fe+} and \ce{Ti+}, that have meaningful inherent opacity; however, their outgassed BSE abundances are relatively small, and thus do not contribute to the models. On the neutral side, some of the strongest absorbers include \ce{Na} and \ce{SiO}, both of which have been noted in previous studies as potentially the most important constituents of silicate atmospheres \citep{Schaefer_2009,Miguel_2011,Ito_2015}. 

Most of \textbf{\ce{Na}} opacity comes from the strong 0.6 \textmu m doublet (see Fig. \ref{fig:Opacity}). For isolated atomic lines, the characteristics of line wings play a significant role in its opacity value. Wings are difficult to model as their exact broadening behaviour is dependant on collisions with the background species. The required theory for arbitrary atmospheres (non-hydrogen) is currently not well understood. In our case, the wings are modelled with an \ce{H2} atmosphere in mind, however, as explained in section \ref{sec:MethodsT}, pressure broadening effects are ignored. To avoid opacity overestimation, the wings are also trimmed at an arbitrary length. In terms of chemistry, \ce{Na} is consistently the leading atmospheric constituent until about 3000 K, at which point it starts to become ionised. Yet, even at lower abundances, its strong opacity keeps it a major shortwave absorber. That said, because \ce{Na} is the most volatile constituent of silicate atmospheres, its dominance may be short lived, as it is expected to be readily lost through escape and condensation away from the sub-stellar point. Evaporation results are outlined in detail in Section \ref{sec:ResultsII}.

\textbf{\ce{Silicon}} is the second most abundant element outgassed from the melt, mostly residing in the form of \ce{SiO}. Its abundance is closely followed by \ce{SiO2}. Contrary to \ce{Na}, \ce{SiO} is much less volatile and its abundance increases with increasing temperature. Its opacity is strong in both shortwave and IR regions. \ce{SiO} shortwave band is strong enough to cause upper atmosphere thermal inversions even if other major shortwave absorbers are excluded. As discussed in Section \ref{sec:ResultsII}, \ce{SiO} long wavelength bands have distinct features that may prove very useful in characterising silicate atmospheres using low resolution spectroscopy. \ce{SiO2} is another major component of silicate atmospheres. It is only recently that the vast \ce{SiO2} molecular list, consisting of 33 billion transitions, has been computed \citep{Owens_2020}. If sustained in gaseous phase, \ce{SiO2} abundance is expected to be nearly $10^{-3}$ in volume mixing ratio for atmospheres above 2500 K, about 2 orders of magnitude lower than the abundance of \ce{SiO}. A significant region of the IR opacity is dominated by \ce{SiO2}. As its bands do not overlap with \ce{SiO}, the combination of the two molecules creates a unique spectral pattern. If \ce{SiO2} is sustained in the atmosphere as a gaseous species, it should be expected to be a major opacity source for all tenable silicate atmospheres.

Many other species are important contributors. Notably, \ce{MgO} and \ce{TiO} are crucial components that further shape the atmospheric structure. \ce{MgO} is expected to be highly abundant for the entire temperature range with its opacity covering both shortwave and IR wavelength regions, acting as a greenhouse continuum. \ce{TiO} is present at the higher temperature range and is one of the dominant shortwave absorbers in the upper atmosphere. Inclusion of both molecules strongly affects the temperature structure and the expected emission spectrum. Some other minor opacities that slightly influence the atmosphere include \ce{Al}, \ce{Ca} and \ce{Mg}, but all can be dismissed if necessary.

In Figure \ref{fig:Opacity} the upper panel displays the total, wavelength and temperature dependant opacity. The contribution for the given temperature is represented by the colour intensity. The fitted curve shows which temperatures have dominant opacity values per each defined wavelength. At $10^{-4}$ bar and atmospheric temperatures above 2000 K, shortwave opacity will be greater than its longer wavelength counterpart. This imbalance will shape the temperature-pressure profile, causing an inversion to occur. This is only true for low pressures. When the pressure increases, molecules become more abundant, thus resulting in greater IR opacity. At high surface temperatures, the thermal structure of silicate atmospheres should be expected to have distinct regions with opposing temperature gradients.


\subsubsection{Temperature-pressure structure}

\begin{figure}[!h]
    \centering
	\includegraphics[width=0.5\textwidth]{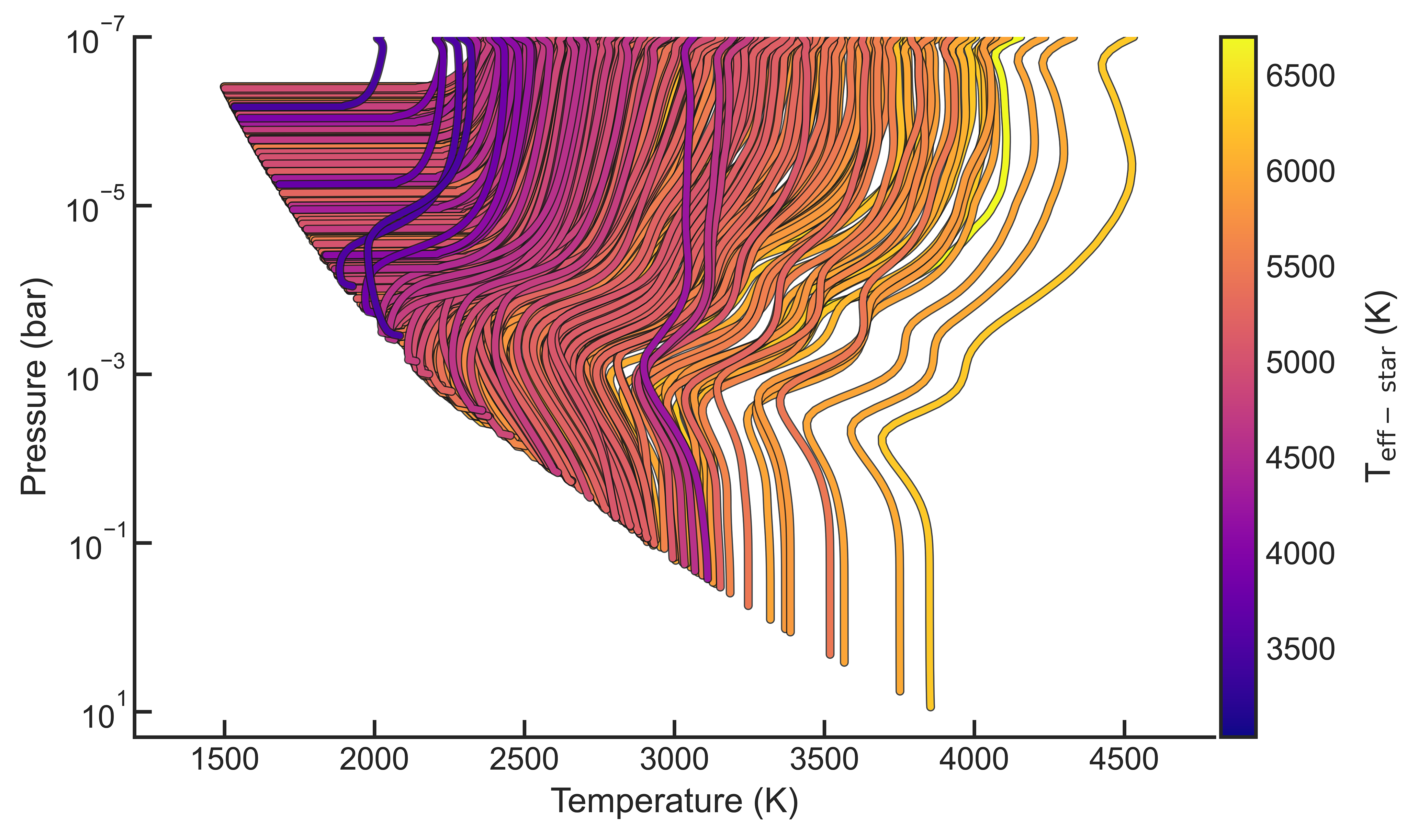}
    \caption{Temperature-pressure profiles of confirmed lava planets with sub-stellar temperatures > 1500 K. Calculated using non-evolved (0\% fractional vaporisation) Bulk Silicate Earth compositions. All profiles are converged to radiative-convective chemical equilibrium using the opacities described in Section \ref{sec:MethodsT}. The colour is indicative of stellar irradiation temperature.
    }
    \label{fig:TP}
\end{figure}

All temperature-pressure (T-P) profiles for lava planets with sub-stellar temperatures > 1500 K are shown in Figure \ref{fig:TP}. At \textbf{1500 K} the atmosphere is extremely tenuous, sustaining just above $10^{-7}$ bar surface pressure. At these pressures the chemistry mainly consists of atomic, shortwave absorbers, resulting in large surface temperature jumps, but otherwise a nearly isothermal profile. Low pressure atmospheres are of little use for observations using low resolution techniques. However, because outgassed pressure scales exponentially with the temperature of the melt, at \textbf{2000 K} it reaches nearly $10^{-3}$ bar. This pressure is enough to allow for more complex molecular composites, such as \ce{SiO}, to exist in abundance. The effects of this are twofold. First, the increased total opacity pushes the photosphere away from the surface, eliminating the sharp temperature discontinuities. Second, at high pressures IR opacity becomes larger than the shortwave opacity, causing inversions to become confined only to the upper atmospheric regions.

At \textbf{2500 K} the IR opacity of silicate atmospheres increases substantially. The surface pressure now reaches $10^{-2}$ bar. Lower regions follow a dry adiabat, with a clear temperature minimum that is followed further up by an inversion. The temperature minimum, otherwise defined as the tropopause, is a good indication of the average photospheric pressure level. Without modelling emission features, one should expect to see strong contributions from the regions just above and below the tropopause. 

Increasing the temperature further will render the atmosphere completely opaque in shortwave and IR wavelengths. Regions below the photosphere are trapped, no longer efficiently radiating. This results in a deep isothermal layer that is inaccessible to observations. While models indicate that at 4000 K the surface pressure would equate to 10 bar, it is unclear whether such strongly irradiated systems could be sustained. It is very likely that the large deposition of irradiation would result in rapid loss of material caused by erosion \citep{Owen_2017,Owen_2019}.

\begin{figure}[!h]
    \centering
	\includegraphics[width=0.5\textwidth]{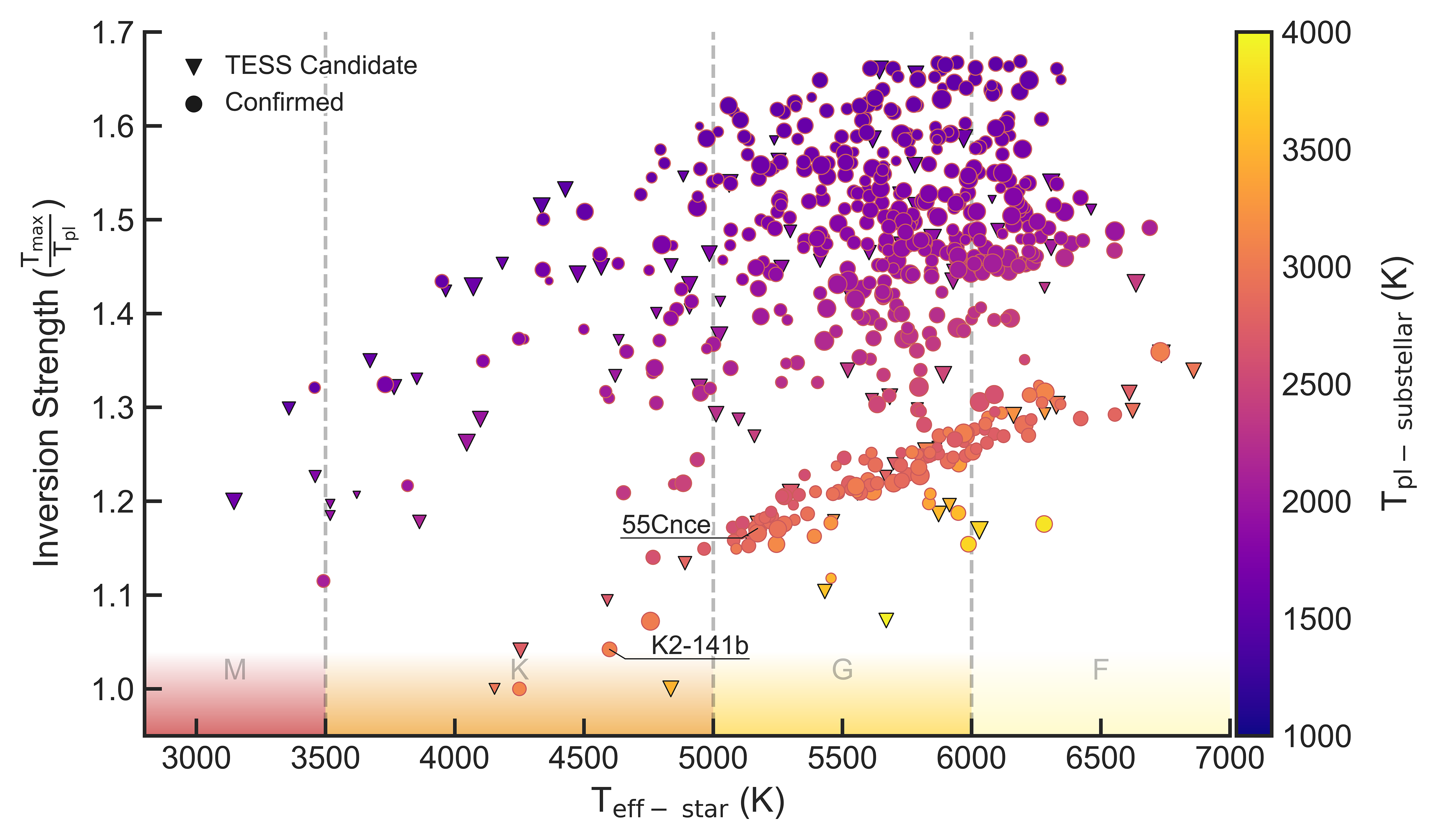}
    \caption{Thermal inversion strength relation with the temperature of the stellar host and the sub-stellar temperature of the planet. Inversion strength is the ratio of the maximum atmospheric temperature with the surface temperature of the planet. In most cases the maximum temperature coincides with the uppermost regions. Inversions become greater with increasing stellar temperature and is anti-correlated with the temperature of the planet. Silicate atmospheres become fully opaque at around 2600 K surface temperature, causing a change in atmospheric structure (isothermal lower layer), indicated via clustered planets separated via a distinct gap.}
    \label{fig:HS}
\end{figure}

Observable emission features depend on the probed thermal structure of the atmosphere \citep{zilinskas_2021}. The shape of the T-P profile is determined by the shortwave/IR absorption of stellar irradiation. The spectral energy distribution coming from hotter stars is much more weighted towards shorter wavelengths. As opposing to cooler stars that peak in IR wavelengths, emitting only a small fraction of energy in the UV/Visible. Figure \ref{fig:HS} demonstrates how the inversion strength scales with temperature of the host star and the sub-stellar surface temperature of planet, which, for BSE, is generally within 50 K of the predicted theoretical value $T_{eq}$ \footnote{Depending on atmospheric composition, the converged surface temperature of the planet may deviate from $T_{eq}$, especially if the planet is strongly irradiated. The effect of this is showcased in Sections \ref{sec:ResultsII} and \ref{sec:ResultsIII}}. Inversion strength is defined by the ratio of the maximum temperature achieved in the model to the temperature of the surface.  In most cases, the maximum temperature coincides with the uppermost regions ($10^{-7}$ - $10^{-5}$ bar), whereas the surface temperature is close to the minimum temperature achieved in the entire atmosphere. We find that inversion strength scales linearly with the temperature of the host star. Thermal inversions will be strongest around hot G and F stars. Planets around cool M or K dwarfs will only experience mild inversions, as evident from purple T-P profiles in Fig. \ref{fig:TP}. Inversion strength is anti-correlated with the surface temperature of the planet. This is attributed not only to the increased chemical complexity and opacity that occurs with greater pressures (larger melt temperatures), but also to thermal dissociation and ionisation of important absorbers, e.g., \ce{Na} or \ce{SiO}. Inclusion of photochemical destruction would likely result in even weaker thermal inversions.

\subsubsection{Observability}

The high sensitivity and large spectral coverage (0.6-28 \textmu m) of the JWST will make characterisation of lava planets achievable. While the tenuous and confined nature of silicate atmospheres render low resolution transmission spectroscopy methods, which rely on atmospheric scale height, unreliable, the scorching dayside temperatures and orbital periods of less than a day make multiple orbit, thermal emission spectroscopy ideal. In addition to outgassed chemical signatures, thermal emission directly probes the temperature structure of the atmosphere, giving us insight into the atmospheric dynamics and occurring radiative processes.

How good a planet is for silicate atmosphere characterisation is determined by several important factors. First, the planet itself has to be of large enough surface temperature to outgas and sustain a substantial atmosphere. Planets with sub-stellar temperatures below 2500 K have pressures lower than $10^{-2}$ bar, which is barely enough to start forming a photosphere. Ideally, a sub-stellar temperature of nearly 3000 K is required to produce an atmosphere that would show large flux deviations from expected blackbody emission. The flux ratio between the planet and the star has to be also large enough to be detectable. This renders a large portion of currently known systems, where the star is too hot or too large in comparison to the planet, unsuitable, even in the IR, where most of the planetary emission comes from. Lastly, the observed system has to be bright enough to have reasonable noise levels, but not so bright that it would saturate the sensors. For MIRI LRS, the saturation limit is expected to be around 5-6 J magnitude \citep{Beichman_2014}. The lower observable limit for the instrument is likely to be around 9-10 J magnitude. MIRI LRS mode itself has spectral capabilities between 5 and 12 \textmu m, making it the ideal mode for characterising the presence of the 9 \textmu m \ce{SiO} feature that has been previously shown to be possibly the best, if not the only characterizable feature of silicate atmospheres \citep{Ito_2015}.

\begin{figure}[!h]
    \centering
	\includegraphics[width=0.5\textwidth]{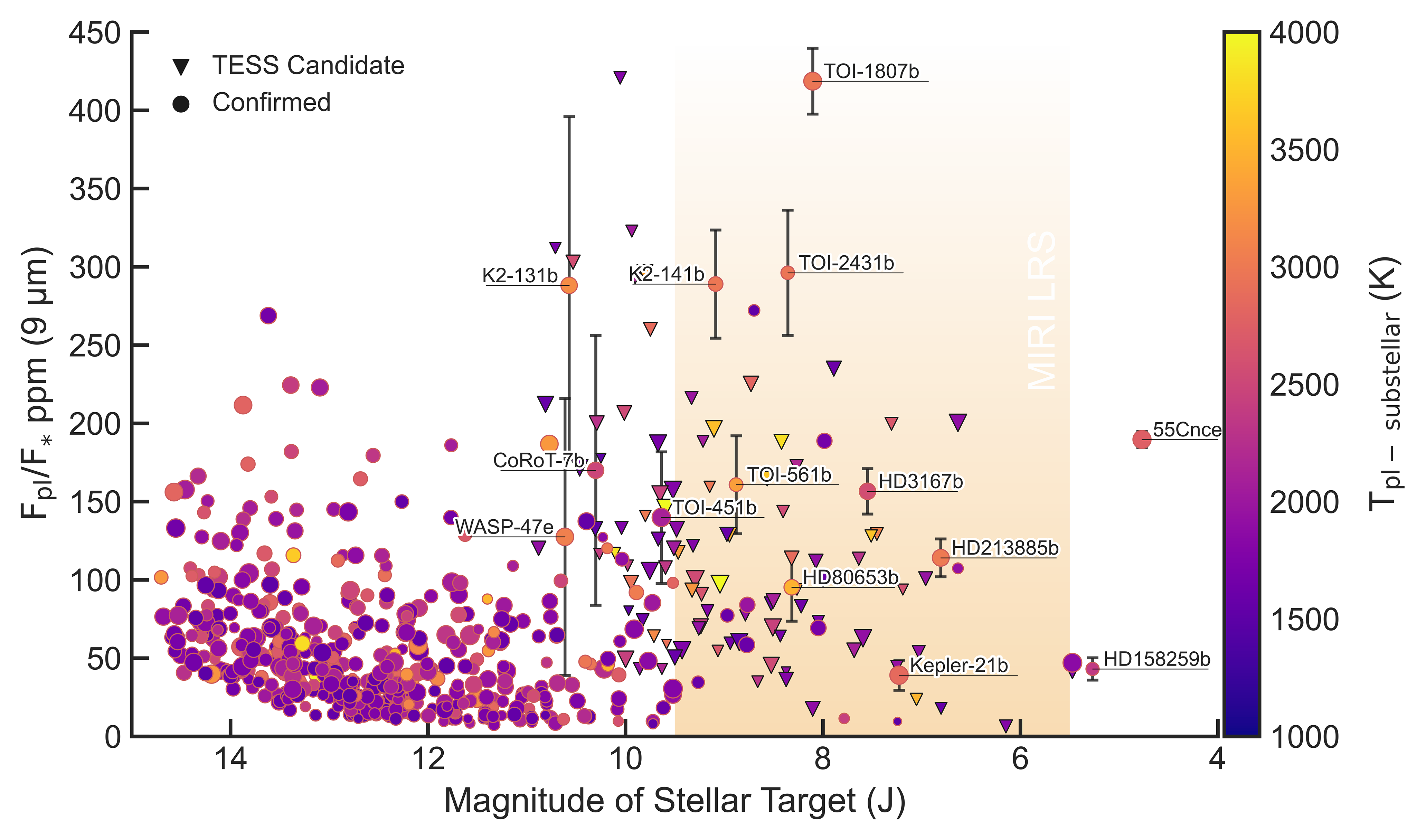}
    \caption{Emission observability of lava planets using the JWST MIRI LRS slitless mode. Flux ratio is computed using radiative-convective models with temperature profiles shown in Fig. \ref{fig:TP}. The flux is given for a wavelength of 9 \textmu m, where the presence of \ce{SiO} is expected to cause a broad emission feature. Error bars are simulated for 20 hours of observations (3 eclipses) and binned to R=10. The highlighted region indicates the optimal magnitude range for the MIRI LRS mode. The colour of each planet denotes its sub-stellar temperature. Ideal targets have high flux ratio, large surface temperature and fall between 9.5 and 5.5 J magnitude.}
    \label{fig:magnitude}
\end{figure}

In Figure \ref{fig:magnitude} we showcase the observability distribution of all modelled lava planets in terms of the above mentioned criteria. Majority of planets are too dim and too cold to be considered for the JWST observations, however, there are some excellent targets, as well as several TESS candidates that should be considered. The standout, \textbf{TOI-1807 b}, is a recently discovered super-Earth orbiting a K dwarf \citep{Hedges_2021}. In the provided sample, it is by far the best candidate for atmospheric characterisation. Its stellar host is comparatively cool and small, resulting in expected planetary flux from the dayside relative to the stellar flux (i.e. eclipse depth) at 9 \textmu m of over 400 ppm. The sub-stellar temperature of the planet is expected to be just over 3000 K which would result in an outgassed atmosphere of $10^{-1}$ bar (BSE melt with \ce{Na}). In addition, with a J magnitude of 8.1, the system falls perfectly within the range of sensitivity for MIRI LRS. Simulated noise levels for 20 hours of observations (3 eclipses, binned to R=10) are just 21 ppm. TOI-1807 b is an exceptional target with no current equal alternatives.

In terms of dayside flux, the closest two planets to TOI-1807 b are \textbf{K2-141 b} and \textbf{TOI-2431 b}\footnote{TOI-2431 b is at the time of writing still a TESS planet candidate but the target is likely to be confirmed soon due to an accepted HST programme (GO 16660, S. Quinn 2021).}. As their sub-stellar temperatures are comparable to TOI-1807 b, both of the planets are also expected to possess substantial silicate atmospheres. With both systems being reasonably bright for the JWST and emission flux reaching nearly 300 ppm, these planets are good candidates for follow up characterisation. K2-141 b is a well studied target \citep[][; Zieba et al. in submitted]{Barrag_2018,Malavolta_2018,Nguyen_2020} and is even approved for JWST GO Cycle 1 programme for over 20 hours with MIRI LRS \citep{Dang_2021} and 12 hours with NIRSpec \citep{Espinoza_2021}. While TOI-2431 b has not been selected for the JWST observations, it may prove to be just as good of a target as K2-141 b for future proposals.

Other notable targets within the brightness limits of MIRI LRS include: a 3300 K super-Earth orbiting a sun-like star TOI-561 b; a similar but cooler, 2300 K super-Earth HD 3167 b and a very bright, 3000 K super-Earth HD 213885 b. HD 80653 b and Kepler-21 b are strongly irradiated planets that fall into the MIRI range; however, their secondary eclipse flux values are below 100 ppm, making atmospheric characterisation much more difficult. While GJ 367 b is a nearby sub-Earth that is also approved for JWST, its equilibrium temperature is lower than 2000 K, resulting in an outgassed atmospheric pressure of < $10^{-4}$ bar. Such, relatively cool, targets are also unlikely to be good candidates for characterisation of silicate atmospheres.

\textbf{55 Cnc e} currently stands as the brightest known super-Earth. The planet has been under extensive investigation, with the current conclusion that it may harbour a thick, volatile-rich atmosphere, possibly dominated by nitrogen, carbon and/or oxygen bearing species \citep{Demory_2016b,Angelo_2017,ZILINSKAS_2020,zilinskas_2021}. Its high surface temperature makes it also likely to have a partially molten crust, which would result in a silicate-rich atmosphere. Currently, there is not enough spectroscopic data to determine the exact composition, however, the planet will be observed in thermal emission using NIRCam and MIRI LRS modes during JWST GO Cycle 1 \citep{HU_2021}. If successful, it would provide spectral coverage between 3.8 and 12 \textmu m, potentially revealing its atmospheric composition and thermal structure. The large brightness of the system may result in oversaturation, partially hindering the recovery of the full spectrum.

Among all of these targets there is a number of yet to be confirmed TESS candidates. As can be seen in Fig. \ref{fig:magnitude}, many of these have expected sub-stellar temperatures of 2500 K or more, have fluxes of over 100 ppm and lie perfectly within the sensitivity range of the MIRI LRS mode.

\begin{figure}
    \centering
	\includegraphics[width=0.5\textwidth]{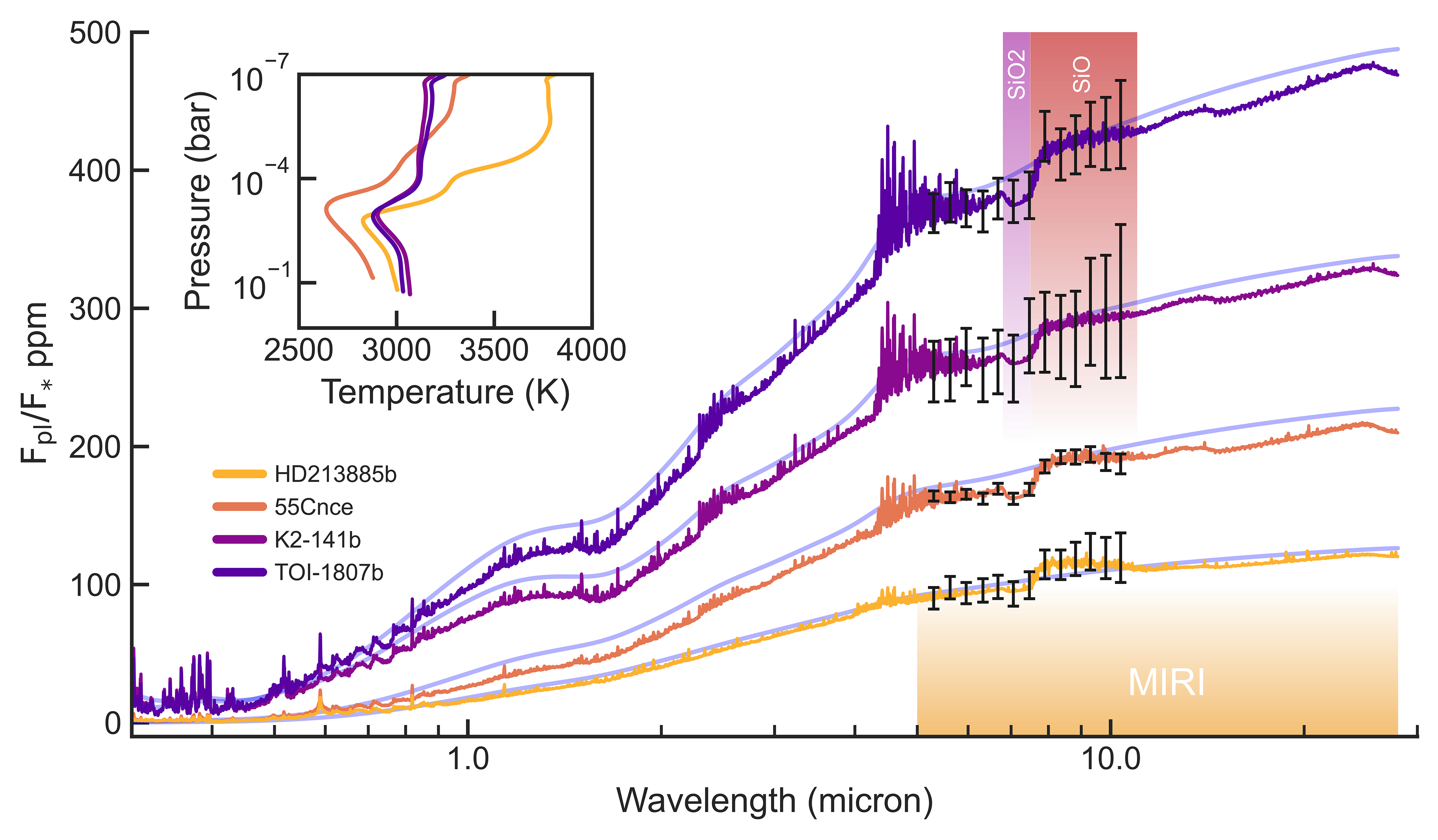}
    \caption{Synthetic emission spectra of four selected lava planets with outgassed silicate atmospheres: HD 213885 b, 55 Cnc e, K2-141 b and TOI-1807 b. The spectra are computed for the entire JWST wavelength range and convolved to a resolution of R=800. The blue curves denote blackbody emission of numerically converged surface temperature. Corresponding temperature profiles for each planet are shown in the inset. Error bars are simulated for the JWST MIRI LRS mode, assuming 20 hours of observations (3 eclipses) and binned to R=10. The indicated MIRI coverage region includes both, low resolution and medium resolution modes. Shaded \ce{SiO} (9 \textmu m) and \ce{SiO2} (7 \textmu m) are likely to be strongest characterizable features of silicate atmospheres. Note that \ce{SiO} also has an additional band between 4 - 5 \textmu m, however, its magnitude is, in most cases, small. The noise at 4 - 5 \textmu m is an artefact of generated synthetic stellar spectra. }
    \label{fig:spectraERror}
\end{figure}

Figure \ref{fig:spectraERror} shows synthetic spectra of four selected planets: HD 213885 b, 55 Cnc e, K2-141 b and TOI-1807 b. While also being good targets, these planets cover a range of different eclipse depth values. The spectra are generated for the range of 0.3 - 28 \textmu m. The wavelengths covered by the MIRI instrument are between 5 - 28 \textmu m. However, MIRI LRS mode already covers the sufficient range (5 - 12 \textmu m) for the major silicate atmosphere features. Thus, the error bars are simulated only for the MIRI LRS slitless mode assuming 20 hours of observations and binned to a resolution of R=10. All four planets have surface temperatures of around 3000 K, with 55 Cnc e being slightly on the cooler side, but they vary strongly in brightness and emission flux. K2-141 b and TOI-1807 b orbit cooler K stars, resulting in weaker thermal inversions, but otherwise similar thermal structure. Their respective temperature profiles are given in the top left inset of the figure.

While these atmospheres are filled with many strong absorbers, the only two relevant species that have broad, visible features are \ce{SiO} and \ce{SiO2}. \ce{SiO} is expected to be the major constituent of all silicate atmospheres. Equilibrium chemistry shows that it is consistently abundant throughout the entire atmosphere, making its opacity relevant even for upper regions, where thermal inversions take place. \ce{SiO} has a strong 9 \textmu m band that is 3.5 \textmu m wide. Because its opacity comes from higher up in the atmosphere, the feature shows up as increased flux. As inversions become weaker with increasing sub-stellar temperature of the planet (see Fig. \ref{fig:HS}), the magnitude of \ce{SiO} features will be diminished for the hottest targets. The bands of \ce{SiO2}, however, do not manifest as emission, but rather as absorption. \ce{SiO2} is mostly confined to the lower, cooler layers of the atmosphere. Its 7 \textmu m \ce{SiO2} band does not overlap with \ce{SiO}, creating a large contrast between the two molecules. The combination could prove to be the best characterisable feature of silicate atmospheres, allowing to probe distinct thermal regions of the atmosphere.

\begin{figure}
    \centering
	\includegraphics[width=0.5\textwidth]{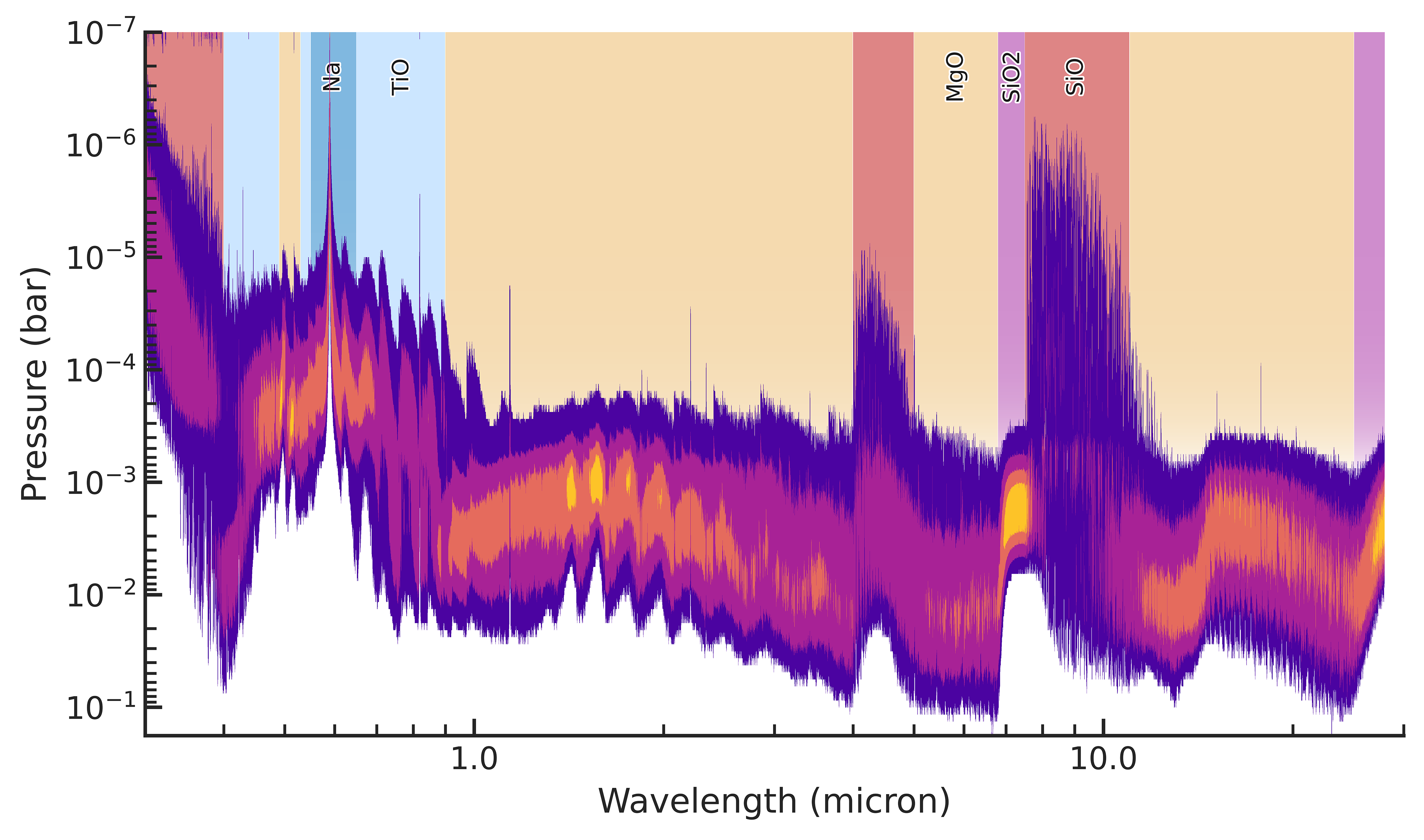}
    \caption{Opacity contribution of an outgassed silicate atmosphere for TOI-1807 b with a non-evolved (0\% fractional vaporisation) Bulk Silicate Earth melt composition. Intensity shaded pressures indicates optically thick regions. All marked species are major contributors of the denoted wavelengths. The corresponding temperature profile and the resulting spectrum are shown in \ref{fig:spectraERror}.}
    \label{fig:contribTOI1807b}
\end{figure}

In Figure \ref{fig:contribTOI1807b} we show the opacity contribution for the atmosphere of TOI-1807 b. Intensity shaded pressures represent the photosphere being probed by emission spectroscopy. The opacity of the 9 \textmu m \ce{SiO} feature is not confined to single atmospheric pressure, but rather becomes non-negligible at pressures as low as $10^{-6}$ bar, extending for several orders of magnitude downwards. On the contrary, \ce{SiO2} is confined to a region that is less than order of magnitude wide. At $10^{-3}$ bar, it mostly coincides with the minimum of the temperature profile, which results in an absorption feature. Going further beyond of the 9 \textmu m \ce{SiO} band, the opacity comes mostly from \ce{MgO}, which acts more or less as a continuum for silicate atmospheres. As with \ce{SiO2}, its presence is also mostly confined to the lower temperature regions. Apart from the small flux change at 13 \textmu m, \ce{MgO} does not have any distinct features. If \ce{MgO} was absent from the atmosphere, beyond 9 \textmu m the spectrum would instead show absorption of \ce{SiO2}, which has a distinctive band between 11 - 16 \textmu m.

While there are additional strong shortwave absorbers, such as \ce{Na} and \ce{TiO}, their spectral features are extremely minor and likely undetectable with low resolution spectroscopy. Even the 0.6 \textmu m \ce{Na} doublet results in only slightly increased flux. In this study we omit pressure broadening effects, which can result in broader \ce{Na} features, but even then its characterisation might prove difficult. Note that this is only true for this particular case. Other outgassed compositions can have increased abundances of shortwave absorbers, resulting in stronger emission features (see Section \ref{sec:ResultsIII}. 


\subsection{Evolved atmospheres}
\label{sec:ResultsII}

As shown by circulation models in \citet{Kite_2016}, planets with sub-stellar temperatures > 2400 K are likely to be `atmosphere-dominated', meaning that the surface melt composition is dragged away from its initial state by atmospheric evolution itself. This is a direct consequence of material being lost from the atmosphere either through escape to space or being carried to the cooler regions and condensing out from the system. For strongly irradiated planets this evolution is rapid and is expected to outpace mass recycling between the interior and the surface melt. 

In the outgassing code, \texttt{MAGMA}, atmospheric evolution is simulated with fractional vaporisation. It is essentially sequential removal of the most volatile constituents until a set percentage of the total mass has been removed from the system. \ce{Na} is the most volatile component and is in all cases rapidly lost. At 3000 K sub-stellar temperature its abundance becomes vanishingly small at less than 10\% removed material. Because it is a major vapour, its removal causes a large drop in atmospheric pressure, as well as decrease in shortwave opacity. Continuing the evolution, abundances of \ce{K} and \ce{Fe} elements slowly decrease, but neither are completely removed when the melt temperature is 3000 K. \ce{Si} is also lost from the system, albeit much more slowly. While volatile species are reduced, the dominance of refractory elements grows. Abundances of \ce{Mg}, \ce{Ti}, \ce{Ca} and \ce{Al} all steadily increase. Appendix \ref{appendixB} contains figure for the evolution of major atmospheric abundances with fractional vaporisation. Readers looking for a much more detailed analysis of fractional vaporisation and possible removal of volatiles are directed to \citet{Schaefer_2004,Schaefer_2009,Kite_2016}.

\begin{figure}[!h]
    \centering
	\includegraphics[width=0.5\textwidth]{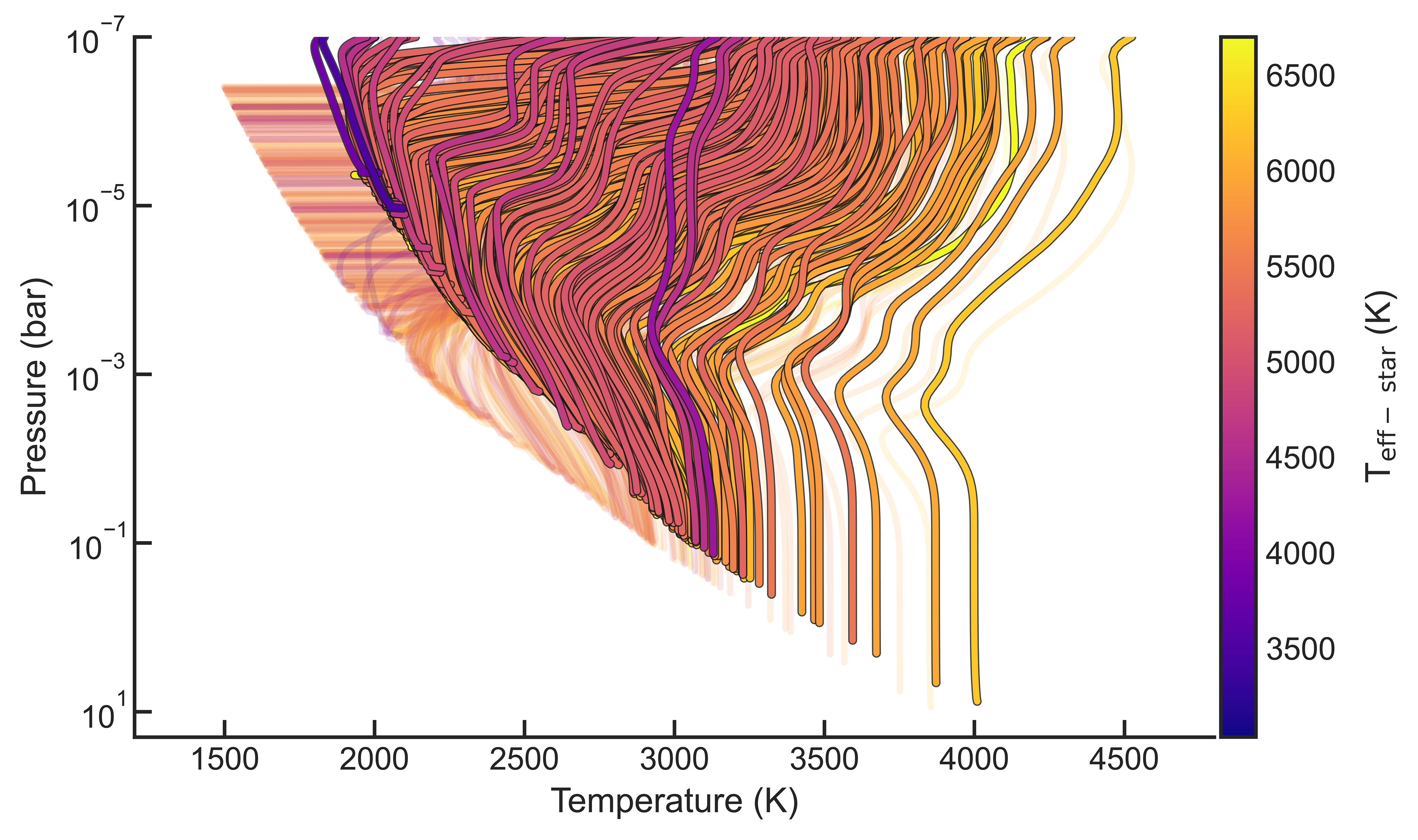}
    \caption{Temperature profiles of evolved silicate atmospheres with sub-stellar temperatures > 2000 K, computed for a Bulk Silicate Earth composition with 60\% fractional vaporisation. All profiles are converged to radiative-convective chemical equilibrium using the opacities described in Section \ref{sec:MethodsT}. Stellar irradiation temperature colour coded. Faint curves indicate temperature profiles for non-evolved cases, (0\% fractional vaporisation) originally in Fig. \ref{fig:TP}.
    }
    \label{fig:TP60}
\end{figure}

Figure \ref{fig:TP60} showcases T-P profiles at 60\% fractional vaporisation, for surface temperatures larger than 2000 K. The faint background curves indicate T-P profiles with no vaporisation (from Fig. \ref{fig:TP}. Depletion of \ce{Na} has the largest effect on the thermal structure, causing the pressure to drop nearly two orders of magnitude at sub-stellar temperatures near 2000 K. At higher temperatures, \ce{Na} is less dominant, hence its removal has a lesser effect on the total vapour pressure. However, in all cases, its shortwave opacity is completely removed, causing thermal inversions to weaken. With removal of volatiles, \ce{MgO} becomes more abundant, causing the photosphere and the temperature minimum to shift upwards to lower pressures.
The surface temperature of the planet is also affected by the changes in opacity, resulting in a greenhouse effect of up to 200 K. The severity of this is greater for hotter planets that are subject to stronger shortwave irradiation.

\begin{figure}[!h]
    \centering
	\includegraphics[width=0.5\textwidth]{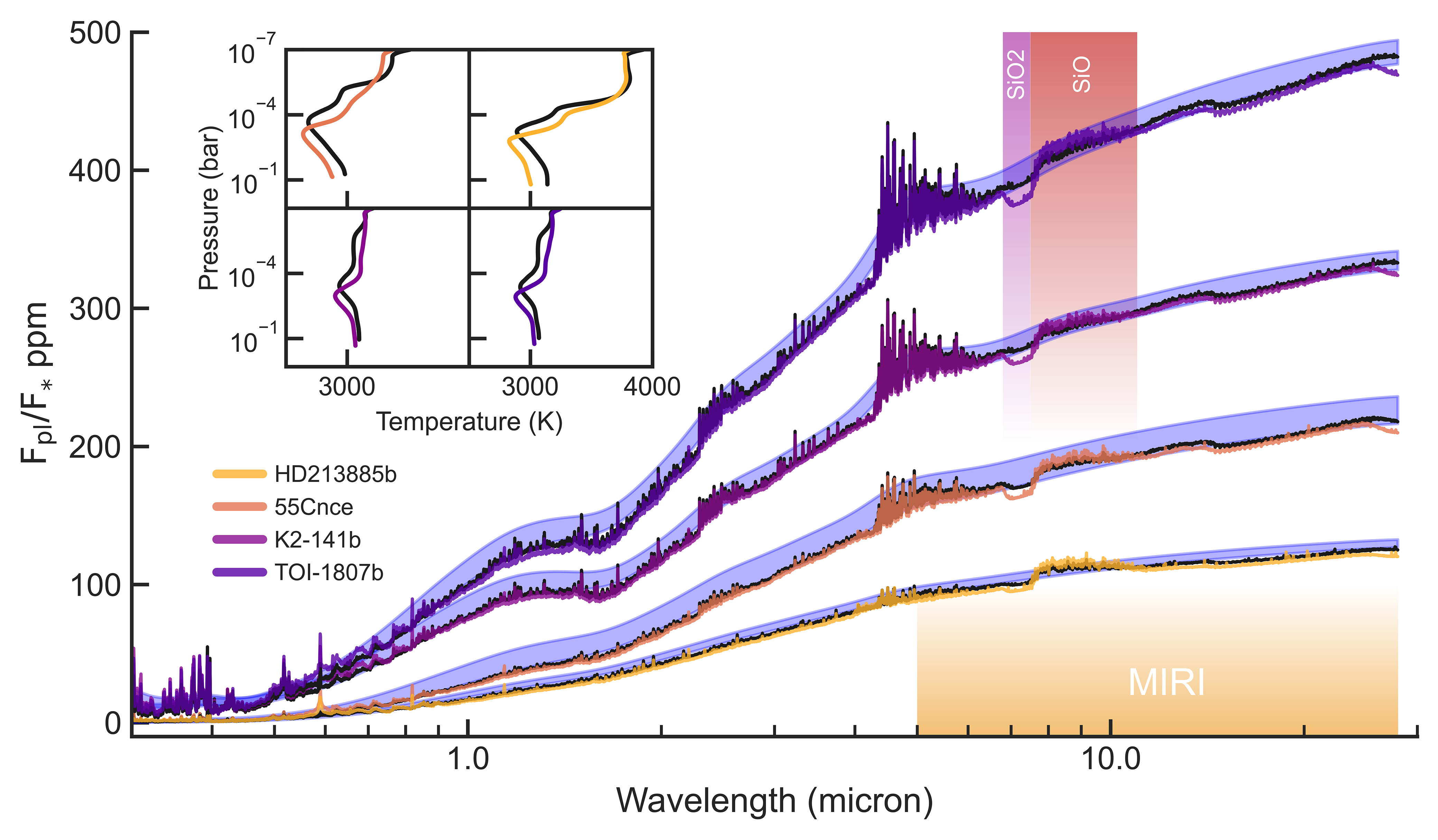}
    \caption{Emission spectra of same four planets as in Fig. \ref{fig:spectraERror}, but now showing the differences caused by fractional vaporisation. Coloured curves indicate atmospheres with no fractional vaporisation, while black curves show atmospheres vaporised to 60\%. All cases are for Bulk Silicate Earth melt composition shown at R=800. Corresponding temperature profiles for each case are shown in the inset. Blue shaded regions now denote blackbody emission boundaries between the converged surface temperature of non-evolved and evolved models.}
    \label{fig:evapSpectra}
\end{figure}

The resulting differences in emission spectra are indicated in Figure \ref{fig:evapSpectra}. The coloured spectra represent the original, non-evolved cases. The black spectra show 60\% fractional vaporisation models. While differences seem small, we find that planets around cooler stars will have their spectral features severely diminished. This is the combined result of the temperature profile shifting closer to an isotherm (reduced inversions), as well as decreased abundances of the absorbers, e.g., \ce{SiO} and \ce{SiO2}. The spectra now coincide closely with the expected lower blackbody emission limit. The effect is slightly lesser for planets around hotter stars, but generally similar. Needless to say, removal of \ce{Na} removes any possible emission from 0.6 \textmu m doublet. Different vaporisation percentages show similar results, with the general trend of spectral features disappearing and the spectra moving closer to blackbody emission. If the effects of vaporisation are prominent on lava worlds, it would make characterisation of their atmospheres with low resolution spectroscopy much more demanding. 

\subsection{Atmospheres of varying melt compositions}
\label{sec:ResultsIII}


The bulk composition of lava planets can be estimated from their measured density \citep{Dorn_2015}, the properties of the specific host star \citep{Santos_2017,Putirka_2019}, or even from polluted white dwarfs \citep{Klein_2011,Putirka_2021}. However, besides inherent degeneracy of interior models complicating the matter, the formation history of the planet can also strongly affect the composition \citep{Bond_2010,Carter_2012}. If the predictive assumption of silicate atmospheres is correct, determining atmospheric abundances from observations could allow for better constraints on melt compositions and possibly interiors. Until then, our best examples come from our own solar system.

\citet{Hart_1986} introduced the term Bulk Silicate Earth, representing the combined composition of the early Earth's crust and mantle, which excludes the iron-rich core. While strongly irradiated rocky worlds may resemble early Earth, currently there is no evidence for it. Thus, besides BSE, we consider several other viable melts that vary in their oxide ratios.

Listed in Table \ref{table:melts} are: BSE, Continental, Oceanic, Komatiite, and Mercury. Continental and Oceanic represent parts of the Earth's upper lithosphere, which formed from the molten silicate mantle. Compared to BSE, both compositions are substantially less dominated by \ce{Mg} oxides, but much richer in \ce{Si}, \ce{Al}, \ce{Na}, and \ce{Ti}. Oceanic contains a large amount of \ce{Ti} and \ce{Ca}, while Continental is more enriched in \ce{K}. It has been argued the Komatiite lava, erupted on the early Earth, might be more representative of molten, more massive Earths \citep{Miguel_2011}. Komatiite is resembled by BSE, but with slightly higher \ce{Si} and lower \ce{Mg} content, while also being much more rich in \ce{Fe}. Mercury melt is uniquely poor in \ce{Fe} and due to its close formation proximity to the Sun, it is essentially void of two most volatile components, \ce{Na} and \ce{K}. Its composition is otherwise similar to Komatiite.

\begin{figure*}[!h]
    \centering
	\includegraphics[width=1.0\textwidth]{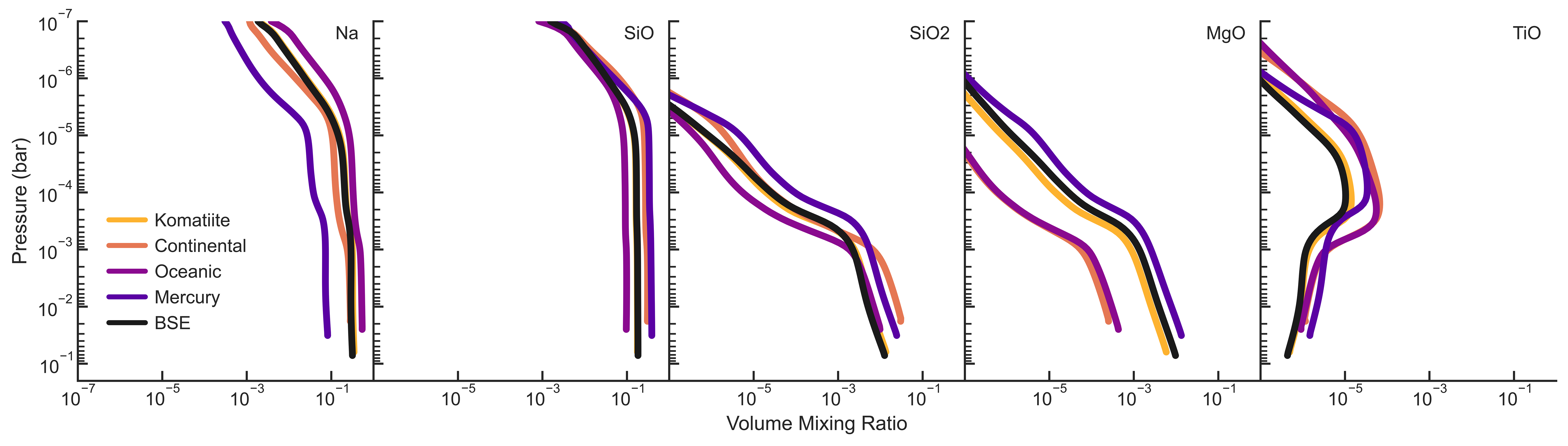}
    \caption{Volume mixing ratio of silicate atmospheres outgassed from Komatiite, Continental, Oceanic, Mercury and BSE melts for 55 Cnc e, with a surface temperatures close to 2700 K. Exact melt compositions are given in Table \ref{table:melts}. The five listed species have the dominant opacities that shape the thermal structure and emission spectrum of the atmosphere. Depending on the exact surface temperature and volatile content, the total outgassed pressure will vary between different melts.}
    \label{fig:chemistryComp}
\end{figure*}

Resulting atmospheric abundances for each of the compositions are indicated in Figure \ref{fig:chemistryComp}. The selected cases are for 55 Cnc e, with sub-stellar temperature ranging between 2600 - 2800 K. The showcased species are all major absorbers of silicate atmospheres.

\textbf{BSE} and \textbf{Komatiite} melts result in nearly identical atmospheric abundances and structure. This is not surprising, since the two melts have very similar volatile reservoir. While Komatiite does have lower \ce{Mg} content, diminishing atmospheric pressure, its higher \ce{Fe} abundance counteracts this effect. In terms of characterisation, the two compositions are essentially identical.

With \textbf{Mercury} having nearly depleted \ce{NaO} and \ce{K2O} in the melt, the results are somewhat similar to the previous compositions. \ce{Na} is so volatile that even a small fraction of it in the melt results in it being a major outgassed species. Partial pressures are now more evenly divided between all different elements (normally dominated by just \ce{Na} or \ce{SiO}). Since this melt is less volatile, it results in slightly lower total pressure.

The two compositions that stand out from the rest are \textbf{Oceanic} and \textbf{Continental} (crustal melts). Differences are caused mostly by their lack of \ce{MgO} and high abundance of \ce{TiO}. Even while the crustal melts are much more abundant in \ce{Na}, diminished \ce{MgO} leads to a substantial pressure drop and much lower surface temperatures. Oceanic composition pressure is still largely dominated by \ce{Na}, whereas \ce{SiO} is leading for Continental. Crustal compositions are uniquely rich in \ce{TiO}, enhancing shortwave absorption in the upper atmosphere. Both compositions, in general, have much greater shortwave opacity, resulting is stronger inversions. Altogether, we find that the variation in melt composition results in moderate thermal structure and atmospheric abundance changes.

\begin{figure}[!h]
    \centering
	\includegraphics[width=0.5\textwidth]{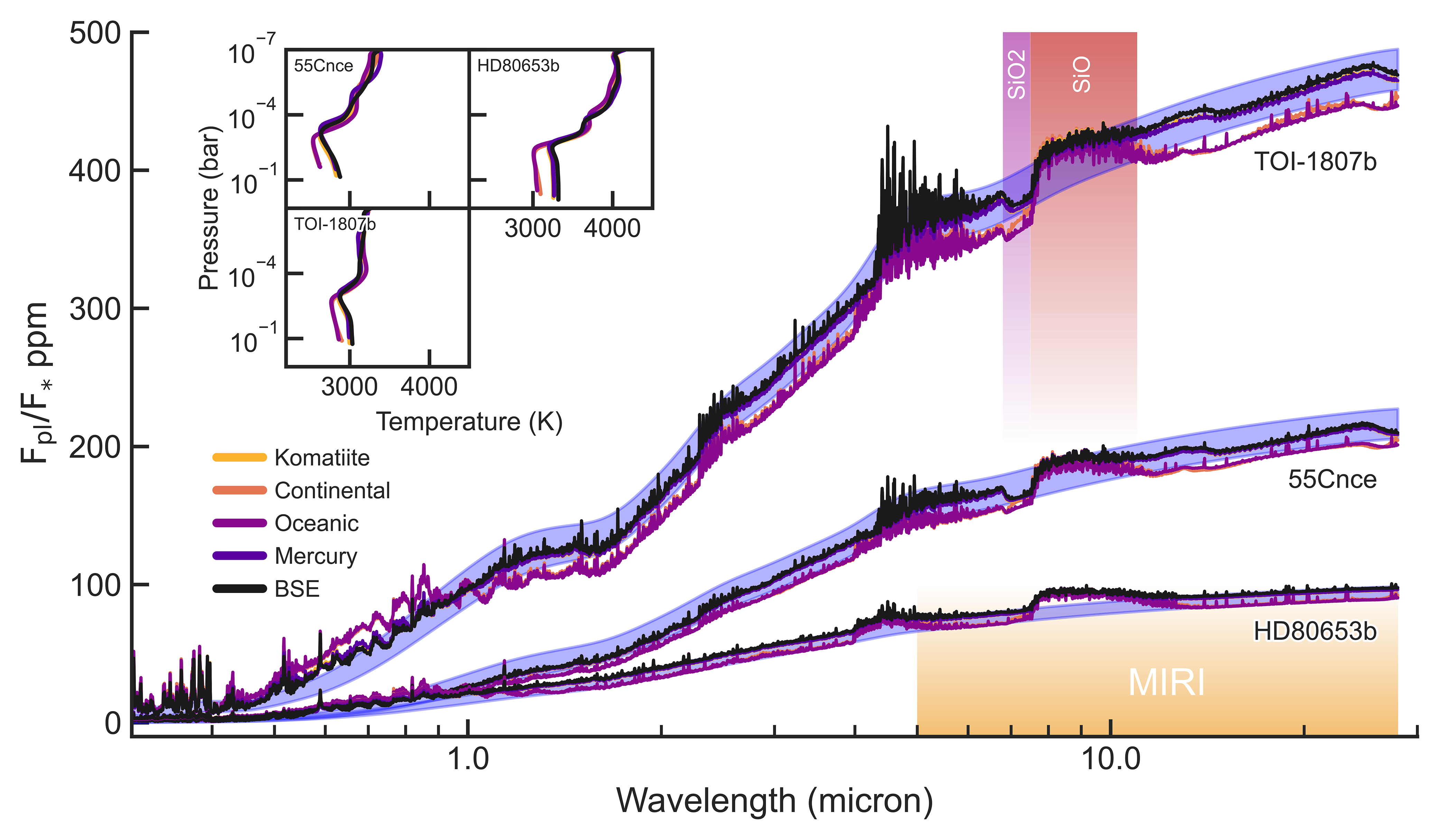}
    \caption{Emission spectra for different melt compositions at R=800. The shown planets are: TOI-1807 b; 55 Cnc e; HD806563 b.
    For each planet different colours indicate varying melt compositions, while the black curves represent Bulk Silicate Earth melts. Corresponding temperature profiles for each of the compositions are show in the inset. The blue shaded region represents variation in surface temperature, which in this case is largest between Bulk Silicate Earth and Oceanic comps.}
    \label{fig:spectraComp}
\end{figure}

The caused spectral differences are given in Figure \ref{fig:spectraComp}. The three displayed planets are TOI-1807 b, 55 Cnc e, and HD 80653 b; picked to cover a range of different stellar and planet surface temperatures. The T-P profiles (shown in the inset) vary only slightly between the bulk compositions (BSE, Komatiite and Mercury), but are differentiated from crustal melts (Continental, Oceanic), which cause a substantial change in the structure of the lower atmosphere. This is reflected in the blackbody emission range, indicated via shaded blue region, which is now much larger than in previous cases (see Fig. \ref{fig:spectraERror} and \ref{fig:evapSpectra}). The differences are smaller for planets orbiting cooler stars, implying that the change in shortwave opacity is the most likely cause in this discrepancy. 

From emission, bulk compositions are likely to be indistinguishable with low resolution spectroscopy. However, outgassed crustal atmospheres have large enough differences to be possibly identifiable. Crustal compositions have stronger inversions and larger \ce{SiO} abundance, which directly results in more contrasting 9 \textmu m \ce{SiO2}/\ce{SiO} features. For bulk atmospheres these features are followed by continuous absorption of \ce{MgO}. But in crust-like atmospheres \ce{MgO} is expected to be severely depleted. Since \ce{SiO2} has a distinct band between 11 - 16 \textmu m, its opacity may replace \ce{MgO}. Its features would show up as absorption, coming from the lower regions of the atmosphere. Detected \ce{SiO2} features over \ce{MgO} would imply that the planetary atmosphere is being outgassed from an Earth-like crust. Such melts are also very enriched in \ce{Ti}, which results in a larger atmospheric abundance of \ce{TiO}. \ce{TiO} is a strong absorber < 0.9 \textmu m. In our models its presence causes a relatively large, continuous increase in flux, which may indeed be observable for the very best targets with the JWST.



\begin{figure}[!h]
    \centering
	\includegraphics[width=0.5\textwidth]{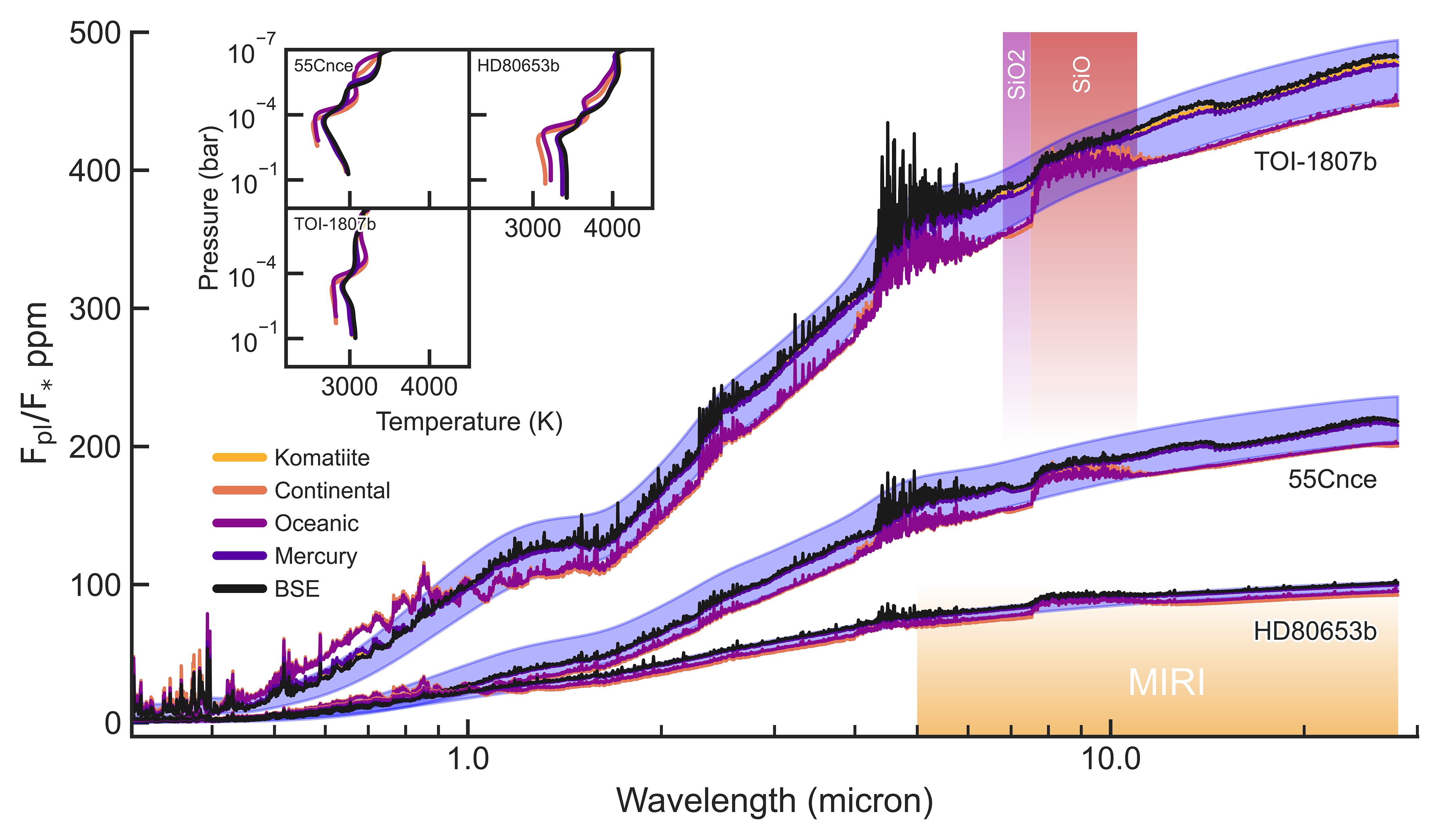}
    \caption{Same as Fig. \ref{fig:spectraComp}, but for 60\% vaporised melts shown at R=800. For each planet, different coloured curves represent different melt compositions. The blue shaded regions represent blackbody emission boundaries between Bulk Silicate Earth and Oceanic compositions. Temperature profiles are given in the inset.}
    \label{fig:spectraCompEvap}
\end{figure}

The situation is similar for evolved atmospheres. Figure \ref{fig:spectraCompEvap} indicates melts with 60\% fractional vaporisation. Removal of volatile inventory causes atmospheric composition to be more dominated by \ce{MgO} and \ce{TiO}, resulting in increased temperatures. Vaporisation causes much larger atmospheric structure difference between the two classes of melts, with the total vapour pressure reduced significantly for crustal compositions. As abundances of \ce{Si} composites drop and inversions become weaker, its features are also diminished. However, unlike for bulk melts, atmospheres produced from crustal melts still have a strong 9 \textmu m \ce{SiO} feature. While \ce{SiO2} is still present in the atmosphere, its detection in evolved atmospheres may be hindered for all plausible compositions. \ce{TiO} features are never removed with vaporisation, as its abundance always grows.

\section{Discussion}
\label{sec:Discussion}
\subsection{Characterising lava worlds}

Current theory predicts that lava planets are either leftover remnant cores of sub-Neptunes, eroded away by intense irradiation, or are simply formed as rocky worlds, without accreting much volatiles \citep{Owen_2017,Lopez_2017}. Whether these planets form in situ or are scattered inwards through gravitational interactions is unclear \citep{Ogihara_2015,Winn_2018}. While there are several observed USP planets inconsistent with volatile-free atmospheres (namely 55 Cnc e \citep{Demory_2016b,Hammond_2017}), most are found to have densities similar to that of Earth \citep{Dai_2019}. It is expected that the permanently irradiated dayside will outgas and sustain a silicate-rich atmosphere \citep{Miguel_2011}. Characterisation of short period planets is an intriguing next step in the exoplanet field. JWST will provide much needed spectral coverage to make this possible. The outstanding issue is what exactly do we look for.

Previous studies have put forward \ce{SiO} as the primary detectable species in silicate atmospheres \citep{Schaefer_2009,Miguel_2011,Ito_2015}. As \citet{Ito_2015} argue, due to strong thermal inversions, the 9 \textmu m \ce{SiO} band would show up as a significant increase in flux coming from the planet. In our models thermal inversions are not as prevalent and mostly confined to just the upper regions. The discrepancy is mainly due to our inclusion of up-to-date opacities while also using more accurate  stellar irradiation and self-consistent radiative-transfer models. However, because \ce{SiO} opacity is so dominant, it becomes significant even in very low pressures, where inversions do occur. Thus we still see the 9 \textmu m \ce{SiO} band as emission, though to a much lesser extent.

The additional inclusion of \ce{SiO2} in our models indicates that it would manifest as absorption bands, possibly surrounding the 9 \textmu m emission of \ce{SiO}. \ce{SiO2} abundance increases significantly towards the lower parts of the atmosphere, which coincide with decreasing atmospheric temperatures. The combination of the two molecules at different atmospheric regions may prove to be unique, identifying features of silicate atmospheres. Detecting both as absorption and emission would allow us to put strong constraints on the temperature gradient. Presence of \ce{SiO2} may also indicate a lack of \ce{MgO}, which can be tied to the composition of the melt. Lack of \ce{SiO2} could also suggest volatile depleted melts and/or possibly strong atmospheric dynamics. 

Overall, we find that the variation of melt compositions and the stage of evolution (volatile material removal through escape or condensation) will impact characterisation. Depending on the melt composition, certain emission features may be strongly enhanced or completely removed. Melts that deviate substantially from the expected norm (crustal compositions) will result in large spectral differences. Variation caused by fractional vaporisation is also non-negligible. While with volatile removal the temperatures in the atmosphere are generally increased, the total pressures are lowered. Shortwave opacity reduction also makes inversions weaker. The combination of all these effects can result in the largest emission features, e.g., 9 \textmu m \ce{SiO} band, being reduced and certain bands entirely removed (\ce{SiO2} absorption). As inversion strength decreases with decreasing stellar host temperature, planets orbiting cooler M or K dwarfs may show weaker emission signs, if any at all. Hotter stars will prove to be much better candidates for observing emission features, however, their larger radii and increased temperatures weaken planetary signal, rendering majority of currently known targets unfavourable.  
 
JWST instruments will allow us to easily distinguish between volatile-rich and silicate atmospheres. However, characterising spectral features in silicate atmospheres may only be possible at longer wavelengths. MIRI LRS instrument covering wavelengths between 5-12 \textmu m is ideal for \ce{SiO} and \ce{SiO2} bands. Its high sensitivity will allow for detection of absorption and emission features in just a few secondary eclipses. For the best possible outcome we recommend focusing on targets with expected sub-stellar temperatures larger than 2500 K, preferably orbiting K stars. Some of the best current targets for secondary eclipse observations with MIRI LRS include: TOI-1807 b, 55 Cnc e, TOI-2431 b, K2-141 b, HD 3167 b, TOI-561 b and HD 213885 b (see Figure \ref{fig:magnitude}). 


\subsection{Melts and interiors}

The two melts that may be observationally distinguishable from the rest are Earth's oceanic and continental crust. The Earth's oceanic crust is magnesium and iron-rich, dense, and relatively thin ($\sim$8 km); efficient recycling through the action of plate tectonics maximises the age of oceanic crust on Earth today to about 200 million years \citep{White_2014}. In contrast, the continental crust is richer in silicon and aluminium, less dense, and thicker ($\sim$40 km). Due to its low density, recycling of crust into the interior is far less efficient, which leads to the fact that parts of the continental crust on Earth today exceed 4 billion years in age \citep{Ricker_2014}. Although the crust of the Earth (oceanic+continental) constitutes only $\sim$0.6\% of the mass of the present-day silicate Earth, it contains between 20\% and 70\% of the highly incompatible\footnote{(In)compatibility is a geochemical term used to describe how easily a minor or trace element is able to replace major elements in a mineral.} trace element budget \citep{Rudnick_1995}. Hence, this explains the elevated \ce{Al2O3}, \ce{CaO}, \ce{Na2O}, \ce{TiO2}, and \ce{K2O} values when comparing the crustal to the other melts in Table \ref{table:melts}. For such crustal melts to co-exist, plate tectonics is likely required \citep{Rudnick_2014}. The presence of Earth-like plate tectonics in turn necessitate the presence of a weaker interior layer on which the plates can move, and liquid water is a likely prerequisite to form such a weak layer \citep[e.g.][]{Mierdel_2007}.

Observed spectral signatures correlating with these compositions could indicate that the planet in question has had a recent history of plate tectonics and hence the presence of liquid water. The fact that lava oceans are most likely well mixed with the mantle \citep{Kite_2016} means that the melting of the crust must have occurred very recently due to, for example, a recent giant impact or the host star entering the red giant phase. Therefore it is unlikely that such signatures will be found on targets that are believed to have a long-lived lava ocean. Alternatively, if such signatures are found then it may be an indication that the upper layers of the melt are chemically evolved in such a way that they resemble crustal composition or that the planet has a mantle with a unique bulk composition. 

\subsection{Limitations of 1D equilibrium models}

There are obvious predictive limitations of 1D equilibrium models. This is especially true for lava planets where the atmosphere is likely to be confined to the permanently irradiated dayside. Away from the sub-stellar point, the sharp drop in temperature will induce an enormous pressure gradient, causing strong horizontal winds towards the nightside \citep{Castan_2011,Kite_2016,Nguyen_2020}. Material removed via winds is bound to condense at cooler regions, settling down onto the surface, where it may or may not be reincorporated into the outgassing cycle via surface and interior transport. In this study we have mimicked this effect to some extent using fractional vaporisation, but this is a very simplistic assumption. For a more accurate assessment of observability, 2D Circulation models combined with simplified outgassed chemistry and radiative transfer may prove to be the reasonable next step.

Use of equilibrium chemistry models also ignores the potential effects of photochemistry and vertical mixing. Photochemical destruction/ionisation of strong absorbers, e.g., \ce{SiO} or \ce{Na}, may cause substantial changes in thermal structure and observability of the specific species. Decrease in shortwave absorption will result in weaker thermal inversions. This effect is apparent with evolved cases, where \ce{Na} is removed from the system while \ce{SiO} abundance is slightly reduced (Fig. \ref{fig:TP60}). Approximations of \ce{SiO} photochemistry done by \citet{Ito_2015} show that only pressures below $10^{-5}$ bar would be strongly affected, however, the significance is unclear without employing full kinetic photochemistry networks. The possibility of photochemical haze formation is also not considered.



\section{Conclusions}
\label{sec:Conclusions}

Characterisation of lava worlds is exciting new frontier to be explored in the imminent future. Decade old studies have predicted formation of silicate-rich atmospheres, outgassed from the molten surface of irradiated planets \citep{Schaefer_2009,Miguel_2011}. With the launch of JWST, the topic has received more interest than ever, including many predictive and observational studies \citep[][; Zieba et al. submitted]{Ito_2015,Kite_2016,Dai_2019,Nguyen_2020,Ito_2021}. Several lava planets have been confirmed for the initial observers programme of JWST, which may lead to first evidence of silicate atmospheres on exoplanets \citep{HU_2021,Brandeker_2021,Dang_2021,Espinoza_2021}.  

As current theory predicts, these atmospheres should be depleted in highly volatile material. Thus characterisation with low resolution spectroscopy is likely to be only feasible via infrared emission coming from the dayside of the planet. In this work we have modelled 1D outgassed equilibrium chemistry consistently with radiative-transfer for all currently confirmed short period rocky exoplanets. We have considered a large number of possible species, including ions, as well as all up to date opacities. Finally, we have assessed observability of the best targets with the MIRI LRS instrument. Our results indicate the following:

   \begin{enumerate}
      \item Thermal inversions may not be as dominant in silicate atmospheres as previously thought. We find that inversions extend all the way to the surface only for planets with sub-stellar temperatures below 2000 K. With larger surface temperature, due to increasing IR dominance over shortwave opacity inversions weaken and become confined to the upper atmospheric regions. Inversion strength also decreases with decreasing stellar temperature, implying that lava planets around M and K dwarfs may show only slightly increased emission flux, if any at all. This severely impacts characterisation of silicate atmospheres.
      
      \item The dominant opacity sources for non-evolved silicate atmospheres with sub-stellar temperatures > 2500 K are \ce{SiO}, \ce{SiO2}, \ce{MgO}, \ce{Na} and \ce{TiO}. Excluding these absorbers from models may result in inaccurate temperature-pressure profiles.
      
      \item The best observable features come from silicon oxides, specifically 7 \textmu m \ce{SiO2} and 9 \textmu m \ce{SiO} bands. \ce{SiO2} is confined to the lower, cooler regions, manifesting as absorption features. \ce{SiO} is highly abundant in the low pressure, inversion dominated regions. While our temperature result in relatively small atmospheric features, we find that for the best targets, observations of only a few eclipses  with MIRI LRS are needed for a detection.

      \item The composition of the melt and possible volatile removal from the system will impact observability. Fractionally evolved atmospheres will have reduced \ce{SiO} and \ce{SiO2} features and will prove much more difficult to characterise. Large flux from \ce{TiO} (< 1 \textmu m) could suggest atmospheres outgassed from melts resembling the Earth's Continental or Oceanic crust. Detection 11-16 \textmu m \ce{SiO2} features imply depletion of \ce{MgO}, which is also characteristic of crust-like compositions. Detection of certain atmospheric abundances may allow us to constrain melt compositions, surface activity and even interior dynamics.
      
      \item The recommended targets for silicate atmosphere characterisation with low resolution infrared spectroscopy are lava planets with sub-stellar temperatures of at least 2500 K that ideally orbit cooler K dwarfs. Some of the currently confirmed planets to keep an eye on include: TOI-1807 b, 55 Cnc e, TOI-2431 b, K2-141 b, HD 3167 b, TOI-561 b, HD 213885 b.  55 Cnc e and K2-141 b are confirmed for the JWST Cycle 1 GO programme. 
      
   \end{enumerate}

We stress the fact that 1D equilibrium models are limited in their predictability. The large temperature contrast between the day and the night side, circulation and interior effects, photochemistry and kinetics are all likely to influence the thermal structure and atmospheric abundances. There is also a need for more accurate modelling of opacities, accounting for pressure broadening effects and the high temperatures of silicate atmospheres.

\begin{acknowledgements}

We thank Renyu Hu for his insightful comments that helped improve the paper. This work has made use of the VALD database, operated at Uppsala University, the Institute of Astronomy RAS in Moscow, and the University of Vienna. 
\\
\\
We are also very grateful to all the authors of the following packages making this work possible:
\texttt{HELIOS-K} \citep{Grimm_2015,Grimm_2021}; \texttt{HELIOS}\citep{Malik_2017,Malik_2019}; \texttt{FASTCHEM} \citep{Stock_2018}; \texttt{MAGMA} \citep{Fegley_1987,Schaefer_2004}; \texttt{petitRADTRANS} \citep{Molliere_2019,Molliere_2020}; \texttt{PANDEXO} \citet{Batalha_2017}; \texttt{numpy} \citep{Harris_2020}; \texttt{matplotlib} \citep{Hunter_2007}; \texttt{seaborn} \citep{Waskom_2021}.
\\
\\
Supplementary material, including star spectra, temperature-pressure profiles, emission spectra or planetary parameters can be downloaded at https://github.com/zmantas/LavaPlanets.

\end{acknowledgements}
%
%

\bibliographystyle{aa}
\bibliography{references}

\begin{appendix} 

\section{Opacity Data}
\label{appendixA}

\begin{table}
 \caption[]{Sources and parameters of opacities used in this work}
 \label{table:opacities}
 \centering
 \begin{tabular}{llll}
    & Source & Line list & Cutting Length (cm$^{-1}$)\\
  \hline
  \hline
  \\
  \ce{Al} & This work$^a$ & Kurucz$^1$ & -\\
  \ce{AlO} & DACE$^b$ \& This work & ATP$^2$ & 100\\
  \ce{Ca} & This work & Kurucz & -\\
  \ce{CaO} & DACE \& This work & VBATHY$^3$ & 100\\
  \ce{Fe} & This Work & Kurucz & -\\
  \ce{K} & This Work & Kurucz & -\\
  \ce{Mg} & This Work & Kurucz & -\\
  \ce{MgO} & DACE \& This work & LiTY$^4$ & 100\\
  \ce{Na} & This Work$^c$ & VALD3$^5$ & Varied (300 max)\\
  \ce{O} & This Work & Kurucz & -\\
  \ce{O2} & This Work & HITRAN$^6$ & 100\\
  \ce{Si} & This Work & Kurucz & -\\
  \ce{SiO} & This Work & Kurucz (shortwave) \& EBJT$^7$ & Varied (300 max)\\
  \ce{SiO2} & DACE & OYT3$^8$ & 100\\
  \ce{Ti} & This Work & Kurucz & -\\
  \ce{TiO} & DACE \& This Work & Toto$^9$ & 100\\
    \\
  \hline  
    \\
  \ce{Al+} & This Work & Kurucz & -\\
  \ce{Ca+} & This Work & Kurucz& -\\
  \ce{Fe+} & This Work & Kurucz & -\\
  \ce{Mg+} & This Work & Kurucz & -\\
  \ce{Na+} & This Work & Kurucz & -\\
  \ce{O+} & This Work & Kurucz & -\\
  \ce{Si+} & This Work & Kurucz & -\\
  \ce{Ti+} & This Work & Kurucz & -\\  
  
  \\
  \hline
  \multicolumn{4}{p{13.4cm}}{\footnotesize$^{a}$ Opacities are computed with \texttt{HELIOS-K} https://github.com/exoclime/HELIOS-K \citep{Grimm_2015,Grimm_2021} ; \footnotesize$^{b}$ DACE database https://dace.unige.ch/;  \footnotesize$^{c}$ \ce{Na} doublet wings computed using the unified line-shape theory of \citet{Rossi_1985,Allard_2007a,Allard_2007b}; \footnotesize$^{1}$ Kurucz \citep{Kurucz_1992}; \footnotesize$^{2}$ ATP \citep{Patrascu_2015}; \footnotesize$^{3}$ VBATHY \citep{Yurchenko_2016}; \footnotesize$^{4}$ LiTY \citep{Li_2019}; \footnotesize$^{5}$ VALD3 \citep{Ryab_2015}; \footnotesize$^{6}$ HITRAN \citep{Gordon_2017,Chubb_2021}; \footnotesize$^{7}$ EBJT \citep{Gordon_2017,Chubb_2021}; \footnotesize$^{8}$ OYT3 \citep{Owens_2020}; \footnotesize$^{8}$ Toto \citep{McKemmish_2019}. We note that we only use opacities of main isotopes. In this work we make use of the DACE \citep{Grimm_2015,Grimm_2021}, ExoMol \citep{Chubb_2021}, HITRAN \citep{Gordon_2017}, Kurucz \citep{Kurucz_1992} and VALD3 \citep{Ryab_2015} opacity databases.}
\\
 \end{tabular}
\end{table}

\FloatBarrier
\onecolumn
\section{Evolution of Atmospheric Abundances}
\label{appendixB}

\begin{figure*}[!h]
    \centering
	\includegraphics[width=0.95\textwidth]{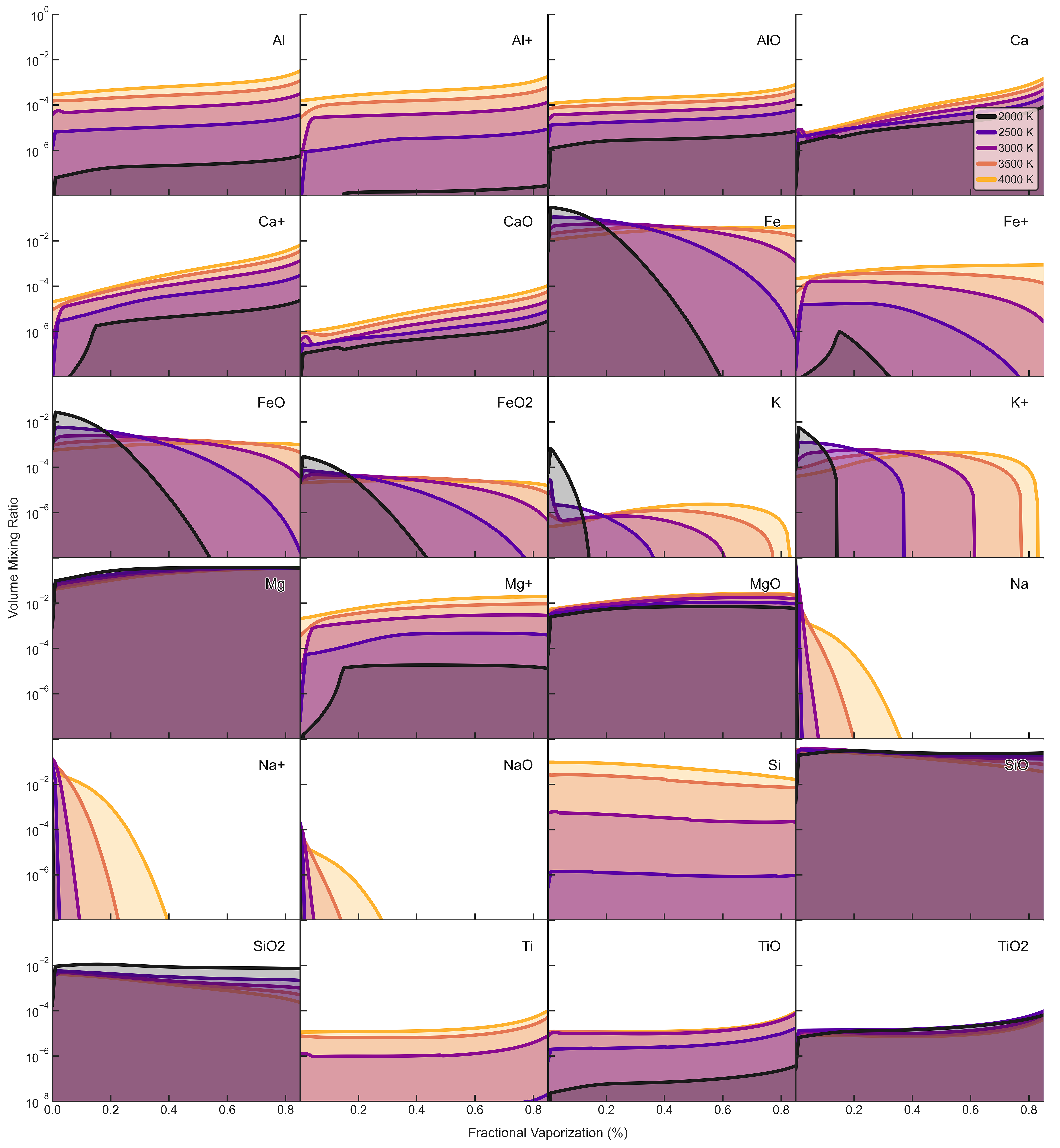}
    \caption{Atmospheric abundance changes with fractional vaporisation (removal of volatile material). Shown for a Bulk Silicate Earth Composition for a range of different surface temperatures. Abundances are taken as averages for all pressures larger than $10^{-5}$ bar, assuming an isothermal temperature profile. Only the most relevant, typically high in abundance, species are shown.}
    \label{fig:appendixAbundances}
\end{figure*}

Figure \ref{fig:appendixAbundances} displays the changes in key atmospheric abundances caused by fractional vaporisation \citep{Schaefer_2004,Schaefer_2009,Miguel_2011,Kite_2016}. The process sequentially removes the most volatile species from the system with \ce{Na}, in all cases, being removed first. Since the outgassed abundances depend on the surface temperature, the vaporisation percentage to remove a certain species is also temperature dependant. Generally, at higher temperatures it takes a larger percentage to remove certain species. With loss of volatiles, refractories become more dominant. This can have substantial effect on the thermal structure of the atmosphere as well as the observable spectrum.

\end{appendix}
\end{document}